\documentclass[twocolumn]{aastex62}

\newcommand{\solm}{M$_{\odot}$\ }

\received{January 1, 2020}
\revised{February 1, 2020}
\accepted{March 30, 2020}
\submitjournal{ApJ}

\shorttitle{Galactic Center S-cluster}
\shortauthors{Ali et al.}

\begin{document}

\title{Kinematic Structure of the Galactic Center S-cluster}

\correspondingauthor{Andreas Eckart}
\email{eckart@ph1.uni-koeln.de}

\author[0000-0002-0786-7307]{Basel Ali}
\affil{ I. Physikalisches Institut der Universit\"at zu K\"oln,
Z\"ulpicher Str. 77, 
D-50937 K\"oln, Germany}

\author{Daria Paul}
\affil{ I. Physikalisches Institut der Universit\"at zu K\"oln,
Z\"ulpicher Str. 77, 
D-50937 K\"oln, Germany}

\author{Andreas Eckart}
\affil{ I. Physikalisches Institut der Universit\"at zu K\"oln,
Z\"ulpicher Str. 77, 
D-50937 K\"oln, Germany}
\affil{Max-Planck-Institut f\"ur Radioastronomie,
Auf dem H\"ugel 69,
D-53121 Bonn, Germany}

\author{Marzieh Parsa}
\affil{ I. Physikalisches Institut der Universit\"at zu K\"oln,
Z\"ulpicher Str. 77, 
D-50937 K\"oln, Germany}
\affil{Max-Planck-Institut f\"ur Radioastronomie,
Auf dem H\"ugel 69,
D-53121 Bonn, Germany}

\author{Michal Zajacek}
\affil{Max-Planck-Institut f\"ur Radioastronomie,
Auf dem H\"ugel 69,
D-53121 Bonn, Germany}
\affil{Center for Theoretical Physics, Polish Academy of Sciences, Al. Lotnikow 32/46, 02-668 Warsaw, Poland}

\author{Florian Pei$\beta$ker}
\affil{ I. Physikalisches Institut der Universit\"at zu K\"oln,
Z\"ulpicher Str. 77, 
D-50937 K\"oln, Germany}

\author{Matthias Subroweit}
\affil{ I. Physikalisches Institut der Universit\"at zu K\"oln,
Z\"ulpicher Str. 77, 
D-50937 K\"oln, Germany}

\author{Monica Valencia-S.}
\affil{ I. Physikalisches Institut der Universit\"at zu K\"oln,
Z\"ulpicher Str. 77, 
D-50937 K\"oln, Germany}

\author{Lauritz Thomkins}
\affil{ I. Physikalisches Institut der Universit\"at zu K\"oln,
Z\"ulpicher Str. 77, 
D-50937 K\"oln, Germany}

\author{Gunther Witzel}
\affil{Max-Planck-Institut f\"ur Radioastronomie,
Auf dem H\"ugel 69,
D-53121 Bonn, Germany}



\begin{abstract}

We present a detailed analysis of the kinematics of 112 stars
that mostly comprise the high velocity S-cluster and orbit the 
super massive black hole SgrA* at the center of the Milky Way.
For 39 of them orbital elements are known, for the remainder we know proper motions.
The distribution of inclinations, and proper motion flight directions 
deviate significantly from a uniform distribution which one expects if the
orientation of the orbits are random.
Across the central arcseconds the S-cluster stars are arranged 
in two almost edge on disks that are located at a position angle approximately $\pm$45$^o$
with respect to the Galactic plane.
The angular momentum vectors for stars in each disk point in both
directions, i.e. the stars in a given disk rotate in opposite ways.
The poles of this structure are located only about 25$^o$ from the line of sight.
This structure may be the result of a resonance process
that started with the formation of the young B-dwarf stars in the cluster
about 6~Myr ago. Alternatively, it indicated the presence of a disturber at 
a distance from the center comparable to the distance of the compact stellar
association IRS13.

\end{abstract}



\section{Introduction}
\label{section:Introduction}

The Galactic Center (GC) stellar cluster harbors a number of stellar associations with 
different ages and potentially different origins.
The luminous 20-30\solm O/WR-stars appear to reside in at least one single 
disk like structure most likely coupled to their formation process
\citep{Levin2003, Yelda2014}.
Their ages have been derived as 6$\pm$2~Myr \citep{Paumard2006}.
The S-cluster consisting of lighter 3.5-20\solm stars, contains the 4 million solar
mass super massive black hole \citep[SgrA*,][]{Gravity2018, Parsa2017}, and appears to be somewhat decoupled 
from the stellar disk at larger radii.
Of K$_s \le$18 stars that reside with separations of less than 1'' 
or those stars that have known semi-major axes of less than 1'' the 
predominant fraction are B-stars. This is especially true for the brightest of the stars
\citep{Habibi2017, Gillessen2017}.

\cite{Gillessen2009} also derive the volume density distribution of the 
the S-cluster B-stars. They find for the 15 stars with a semi-major axes of less than
0.5'' in projection a three-dimensional power law slope of -1.1$\pm$0.3. 
This appears to be marginally larger than the slope derived for a more spread out 
cluster population of B-stars, implying that the S-stars form a distinct possibly
cusp like component.

A detailed near-infrared spectroscopic study of the S-stars 
\citep{Habibi2017, Martins2008, Ghez2003}
shows that these stars are most likely high surface gravity (dwarf) stars
The authors' analysis reveals
an effective temperature of 21000-28500 K, a rotational velocity of 60-170 km/s,
 and a surface gravity of log~{$g$}=4.1-4.2. These properties are characteristic for stars of spectral type B0-B3V
with masses between  8\solm and 14\solm.
Their age is estimated to be less than 15 Myr.
For the early B dwarf (B0–B2.5 V) star S2 \citep{Martins2008} the age is estimated to be 
6.6$^{+3.4}_{-4.7}$Myr. This compares well with the 
age of the clockwise rotating disk of young stars in the GC. 
\cite{Habibi2017} conclude that the low ages for the high velocity stars favor a scenario in which they
formed in a local disk rather than in field binaries subjected to binary disruption and stellar scattering.

The stars in Galactic Bulges or central stellar clusters often show peculiar kinematic
arrangements.
From theory (e.g. \citealt{Contopoulos1988}),
observations of external galaxies, and the Milky Way it has become evident
that boxy and peanut shaped stellar orbits have a significant influence on the appearance 
of galactic bulges.
Perturbations in the vertical direction lead to orbits with boxy appearance
\citep{Chavez2017}.
\cite{Hernquist1992} also described boxy and disk like appearances as possible structures
in post-merger bulges.
\cite{Quillen1997} discovered boxy and peanut-shaped bulges in highly inclined galaxies.
\cite{Quillen2014} 
present a simple resonant Hamiltonian model for the vertical response of a stellar disk to the
growth of a bar perturbation. As the perturbation grows the stars become trapped in vertical Inner
Lindblad resonances and are lifted into higher amplitude orbits. The vertical structure of a boxy and
peanut shaped bulge as a function of radius and azimuthal angle in the galaxy plane can be predicted
from the strength and speed of the bar perturbation and the derivatives of the gravitational potential.
This model predicts that stars on the outer side of the resonance are lifted higher than stars on the inner
side, offering an explanation for the sharp outer edge of the boxy/peanut.

The Milky Way is a barred galaxy whose central bulge has a box/peanut shape and
consists of multiple stellar populations with different orbit distributions
(e.g. \citealt{Gerhard2016}).
Infrared observations revealed
that the Milky Way (MW) bulge shows a boxy/peanut or X-shaped bulge.
Simulations indicate that about 20 per cent of the mass of
the Milky Way bar is associated with the shape \citep{Abbott2017}.

While these structures are associated with resonances linked to a bar or central cluster 
potential, they can also be the result of a perturbation due to an interacting mass.
\cite{Gualandris2009} study the short- and long-term effects of an 
intermediate mass black hole (IMBH) on the orbits of stars
bound to the super-massive black hole (SMBH) at the center of the Milky Way.
The authors consider 19 stars in the S-star cluster and a SMBH mass between 
400 and 4000\solm and a distance from SgrA*
between 0.3-30~mpc. They find that for the more massive perturbers
the orbital elements of the S-stars could experience changes at the level 
of about 1\% in just a few years.  On time scales of 1 Myr or longer,
the IMBH efficiently randomizes the eccentricities and orbital inclinations 
of the S-stars. 
These results support on the one hand that relatively short scale 
response of the S-stars to a nearby perturbation can occur. 
On the other hand the orbits are clearly not fully  randomized implied that
a recent perturbation by massive IMBH within the S-cluster can be excluded.
Resonances could occur however if a perturber is located outside the S-cluster.

In section \ref{section:Observations} we present the observations and data reduction.
In the discussion in section \ref{section:discussion}
we first show in \ref{section:histo}.
histograms and visualizations that highlight our observational results.
In section \ref{subsection:dynamics}
we discuss our findings in terms of stellar dynamical considerations.
A summary and conclusions are given in section \ref{sec:summary}.
Finally, in section \ref{sec:enhanced} we describe the three
supplemented enhanced graphics that
show the projected orbital arrangements in motion.

\section{Observations and data reduction}
\label{section:Observations}

The positions of the S-stars are calculated from the
AO-assisted imaging data of the GC from 2002 to 2015 taken
by the NAOS-CONICA (NACO) instrument installed at the
fourth (from 2001 to 2013) and then the first (from 2014 on)
unit telescope of the Very Large Telescope 
(VLT)\footnote{ProgramIDs: 60.A-9026(A), 713-0078(A), 073.B-0775(A), 073.B-0085(E),
073.B-0085(F), 077.B-0552(A), 273.B.5023(C), 073-B-0085(I), 077.B-0014(C),
077.B-0014(D), 077.B-0014(F), 078.B-0136(A), 179.B-0261(A), 179.B-0261(H),
179.B-0261(L), 179.B-0261(M), 179.B-0261(T), 179.B-0261(N), 179.B-0261(U),
178.B-0261(W), 183.B-0100(G), 183.B-0100(D), 183.B-0100(I), 183.B-0100(J),
183.B-0100(T), 183.B-0100(U), 183.B-0100(V), 087.B-0017(A), 089.B-0145(A),
091.B-0183(A), 095.B-0003(A), 081.B-0648(A), 091.B-0172(A).}.
The K$_s$-band (2.18 $\mu$m) images obtained by the S13 camera (with
13 mas pix$^{-1}$ scale) and the S27 camera of NACO (with
27 mas pix$^{-1}$ scale) are used. The AO guide star is IRS7 with
K$_s$ = 6.5-7.0 mag located at about 5.5'' north of Sgr A*. The
data reduction consists of the standard steps like flat-fielding, sky
subtraction, and bad-pixel correction. A cross-correlation
algorithm is used to align the dithered exposures. We use the
27 mas pix$^{-1}$ scale images to measure the position of the SiO
maser stars IRS9, IRS10EE, IRS12N, IRS15NE, IRS17,
IRS19NW, IRS28, and SiO-15 \citep{Menten1997, Reid2003, Reid2007}, which were needed for finding the
connection of the NACO NIR data and the radio reference
frame. In order to measure the position of the S-stars, the 
Lucy-Richardson deconvolution algorithm is used to resolve the
sources in the 13 mas pix$^{-1}$ scale images. For each epoch we
included all available K$_s$-band frames of the GC stellar cluster
that were taken with a close to diffraction-limited AO
correction and showed Sgr A* flaring. We use the reduced
data presented by \cite{Witzel2012}, Table~2, 2003 to mid-2010, 
and \cite{Eckart2013}, Table~1, and \cite{Shahzamanian2015}, 
Table~1, 2002-2012. 
We supplemented additional imaging data for observing epochs in
2016, 17, 18 for all sources and further 2019 data for the sources
S62, S29, S19, S42, S38, S60.
For the stars S2 and S38 we also used, the positions
published by \cite{Boehle2016} for the years 1995-2010 and
2004-2013, respectively.
As described by \cite{Parsa2017} 
(and following the approach by \cite{Gillessen2009})
the data were added by applying a 
constant linear positional shift between the two data sets.
In addition, we took into account the mean difference between the proper
motions of the VLT and Keck coordinate systems. These differences become 
evident, e.g. in Tab.1 in \cite{Gillessen2017} (see also \cite{Boehle2016}).

\begin{figure*}[htbp!]
	\centering
	\includegraphics[width=\textwidth]{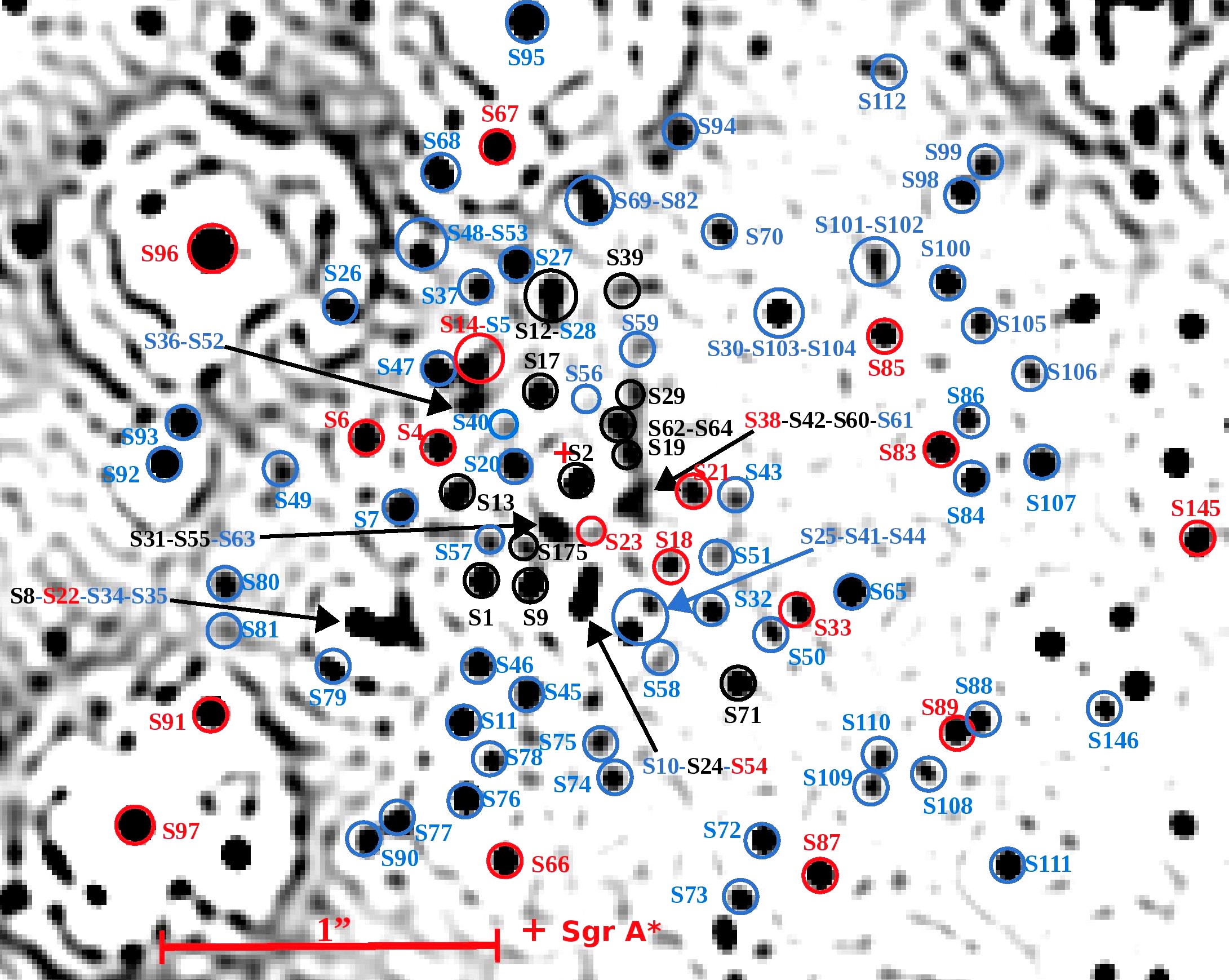}
	\caption{Map of the region showing the S-cluster and some 
neighboring stars. East is to the left, north is up.
We included their nomenclature and encircled 2 or 3 stars
if they happen to be close together at the epoch of the image.
The image was obtained by NACO at the VLT in early 2018.
SgrA*, the counterpart of the super massive black hole, is 
located at the position of the red cross.
Stars encircled by red and black lines belong to the corresponding
disk systems described in section~\ref{section:discussion}.
For all of these stars orbital elements are known.
Blue circles mark the stars for which we only have
short linear sections of their orbits.
}
	\label{fig:map}
\end{figure*}

The selected objects comprise all stars brighter than K$_s$=18.0
that are detectable at all epochs and show no signs of being 
severely confused with other stars of the cluster for most epochs
\citep[see also discussion by][]{Eckart2013, sabha2012}.
An overview image is shown in Fig.\ref{fig:map}.
The positional results were verified by using stars S7, S10, S26, S30, and S65 as
references since these object have almost straight flight paths with 
no detectable  curvature.
For the stars with orbital sections that are short or show no
curvature we fitted a straight line to the flight path.

\cite{Plewa2015} find from the average velocity differences
in radial and tangential directions that the infrared reference
frame shows neither pumping nor rotation relative
to the radio system to within $\sim$7.0 $\mu$as yr$^{-1}$ arcsec$^{-1}$. 
Over 20 years this amounts to an upper limit of 
about 0.14~arcsec, i.e.,
typically to 0.1-0.2~mas across the central 1~arcsec diameter
cluster of high-velocity stars.
Hence, verifying the positional measurements using stars with 
straight flight paths leaves us with an uncertainty of less than
0.5~mas for the 13 mas pix$^{-1}$ scale images.

\begin{figure*}
	\centering
	\includegraphics[width=18cm]{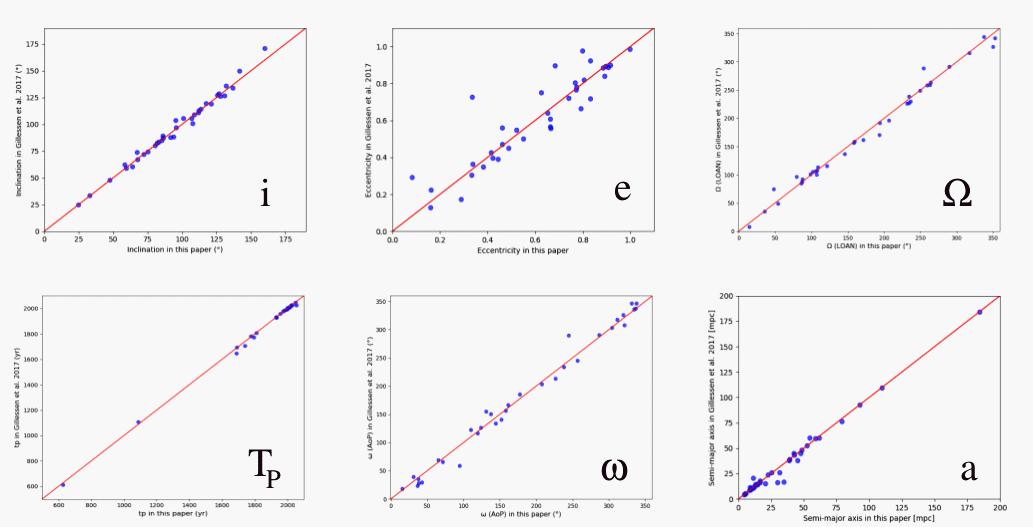}
	\caption{
A comparison between the orbital elements listed by \cite{Gillessen2017}
and in this paper.
}
	\label{fig:calcincs1}
\end{figure*}

In addition to the positional measurements that cover substantially
sections of the curved orbits as made use of the time variable
radial velocities and their uncertainties as  presented in Fig.8 by
\cite{Gillessen2017}
\footnote{This covers: S1, S2, S4, S8, S9, S12, S13, S14,
S17, S18, S19, S21, S24, S31, S38, S54.}.
This includes the radial velocity
data for S2 from the AO-assisted field spectrometer
SINFONI installed on the fourth unit telescope of the VLT and
taken from \cite{Gillessen2009}. The radial velocity
measurements used for S38 are from \cite{Boehle2016}.

\begin{figure}
	\centering
	\includegraphics[width=\columnwidth]{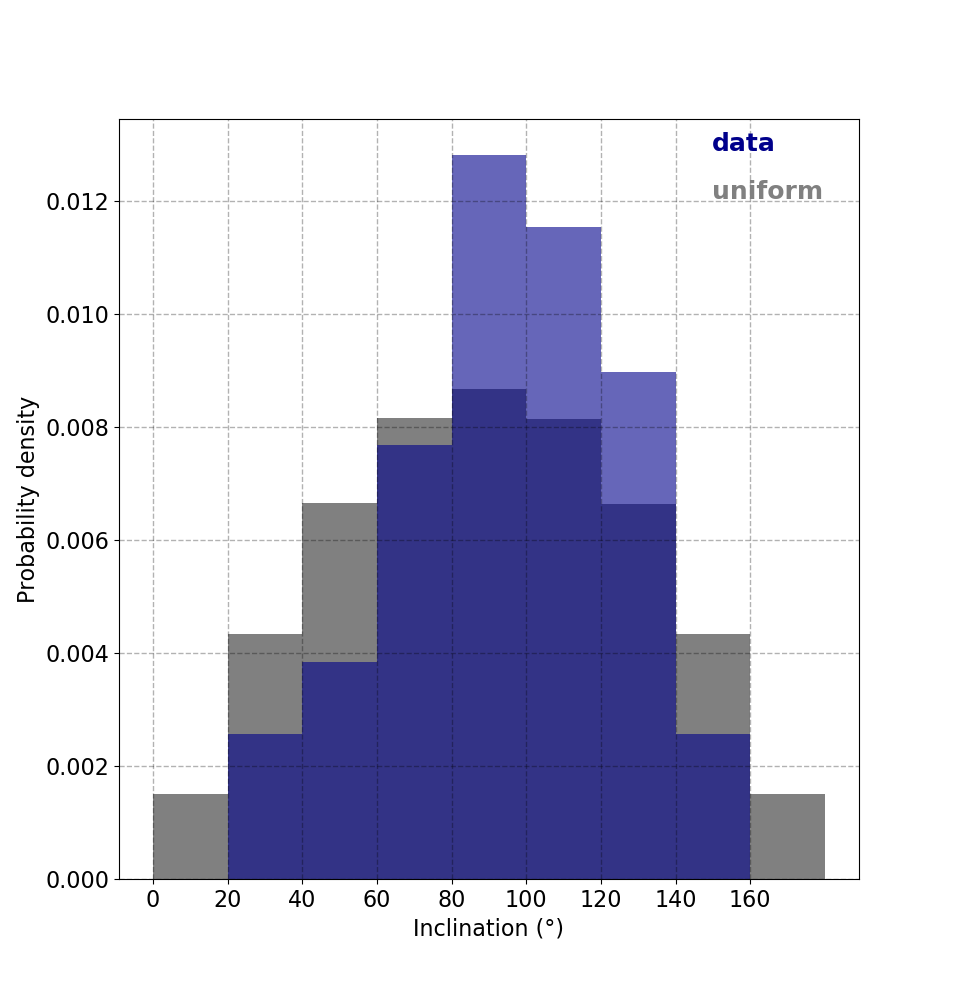}
	\caption{
A comparison between the measured distribution of orbital inclinations
and the expected sin(i) distribution.
}
	\label{fig:calcincs2}
\end{figure}

For the central stars that have larger orbital sections measured 
we modeled the Newtonian stellar orbits 
by integrating the equation of motion using
the fourth-order Runge–Kutta method with twelve or six initial
parameters respectively (i.e., the positions and velocities in
three dimensions).
To determine the six orbital elements a corresponding number of
observables  must be provided. These are the projected positions
$\alpha$, $\delta$, the proper motions $v_{\alpha}$, $v_{\delta}$
the radial velocity  $v_{z}$, and the projected orbital
acceleration. However, higher order moments of the latter two quantities
can also be used as replacements or in support.
The results compare favorably with those of 
fitting routine starting by solving Kepler's equation, which can be
done using the iterative Newtonian method.  This optimization
method is implemented in Python in the Scipy package under the name 
(Sequential Least SQuares Programming) SLSQP.
The optimized results along with boundaries on each of the
elements are then used for bootstrap resampling to get error estimations.
We used a fixed central black hole mass of 4.3$\times$10$^6$\solm
at a distance of 8.3~kpc \citep{Parsa2017, Gillessen2017}.
The results are listed in the appended Tab.\ref{elementsblack} and Tab.\ref{elementsred}.

However, we point out that      
ambiguities in the inclinations of the orbits due to missing
radial velocity information do not affect our prime observables
as used in section~\ref{section:orbitssky}.
These are the directions of the semi-major axes of the sky 
projected orbits and the projected true (i.e., three dimensional) 
semi-major axes of the orbits.
These quantities are listed in the appended Tab.\ref{pablack} and Tab.\ref{pared}.
In total we analyzed 105 S-cluster members and 7 sources 
(S66, S67, S83, S87, S91, S96, S97) that belong to the
the clockwise rotating stellar disk (CWD) of He-stars \citep{Levin2003, Paumard2006}.
This results in our case in 39 stars with orbital elements; for the
remaining stars we fitted straight trajectories.
These compare well within the uncertainties with the parameters derived
for 40 stars by \cite{Gillessen2017}. 
For a discussion of the organization of S-cluster sources see also \cite{Yelda2014}.
For the remaining stars we just fit a straight line to obtain their 
proper motion speed and direction. Their inclusion in the presented discussion
awaits the determination of orbital elements. For completeness we list the
kinematic properties of these stars in the appended tables
(appended Tab.\ref{linear1} and Tab.\ref{linear2}).

\section{Discussion}
\label{section:discussion}

A close inspection of the orbital parameters showed that the stars in the
central arcseconds are arranged in two orthogonal disks.
There are three observational facts that support this finding:

$\bullet$
The distribution of inclinations clustering around 90$^o$.
This shows that stellar orbits are seen preferentially edge on.

$\bullet$
The distribution of semi-major axes of the projected ellipses in the sky
show that the stars populating the disks can indeed be separated into two 
groups.

$\bullet$
The observation of accumulation of orbits that appear face on or edge on
from certain directions
shows the presence of two orthogonal stellar disks.

In the following we describe these findings in more detail and then highlight
present stellar dynamical concepts that may explain the phenomenon.

\subsection{Histograms and Visualizations}
\label{section:histo}

\subsubsection{Orbital inclinations}

In Fig.\ref{fig:calcincs1} we show that the inclinations derived by us and those provided by 
\cite{Gillessen2017} are in very good agreement.
The same can be said for all of the other orbital elements shown in Fig.\ref{fig:calcincs1}.
In Fig.\ref{fig:calcincs2} we show the distribution of all 39 stars with orbital fits
in comparison to a $sin(i)$  distribution as one might have expected for a             
fully uniformly randomized scenario.
Here, $sin(i)$ refers to the expected shape of uniformly distributed inclination
angles and not the trigonometric sine function of the angle.
This ideal shape is also referred to as the Gilbert-sine distribution \citep{Gilbert1895}.
In Fig.\ref{fig:calcincs2} both distributions are normalized to an integral value of unity.
Compared to the $sin(i)$ distribution the measured distribution shows a deficit of stars 
with inclinations in the intervals 0$^o$ to 20$^o$ and 160$^o$ to  180$^o$. 
It also has a full width at half power around only 80$^o$ although one would expect a width of about 100$^o$
for a $sin(i)$ distribution. In addition the measured distribution shows an excess of stars around
inclinations of 80$^o$ to 140$^o$.
The preference for high inclinations can also not be due to a field of view
effect due to the small size of the S-cluster within the large galactic center 
stellar cluster (see Appendix \ref{appendixfield}).
Also, biases for the orbital elements due to incomplete orbital coverage
are not important for the analysis of our problem  (see Appendix \ref{appendixbias}).
Hence, this comparison shows that in the set of 39 S-cluster stars edge on orbits are preferred.

\subsubsection{Distribution of orbits in space}
\label{section:orbitsspace}

\begin{figure*}[htbp!]
	\centering
	\includegraphics[width=0.9\textwidth]{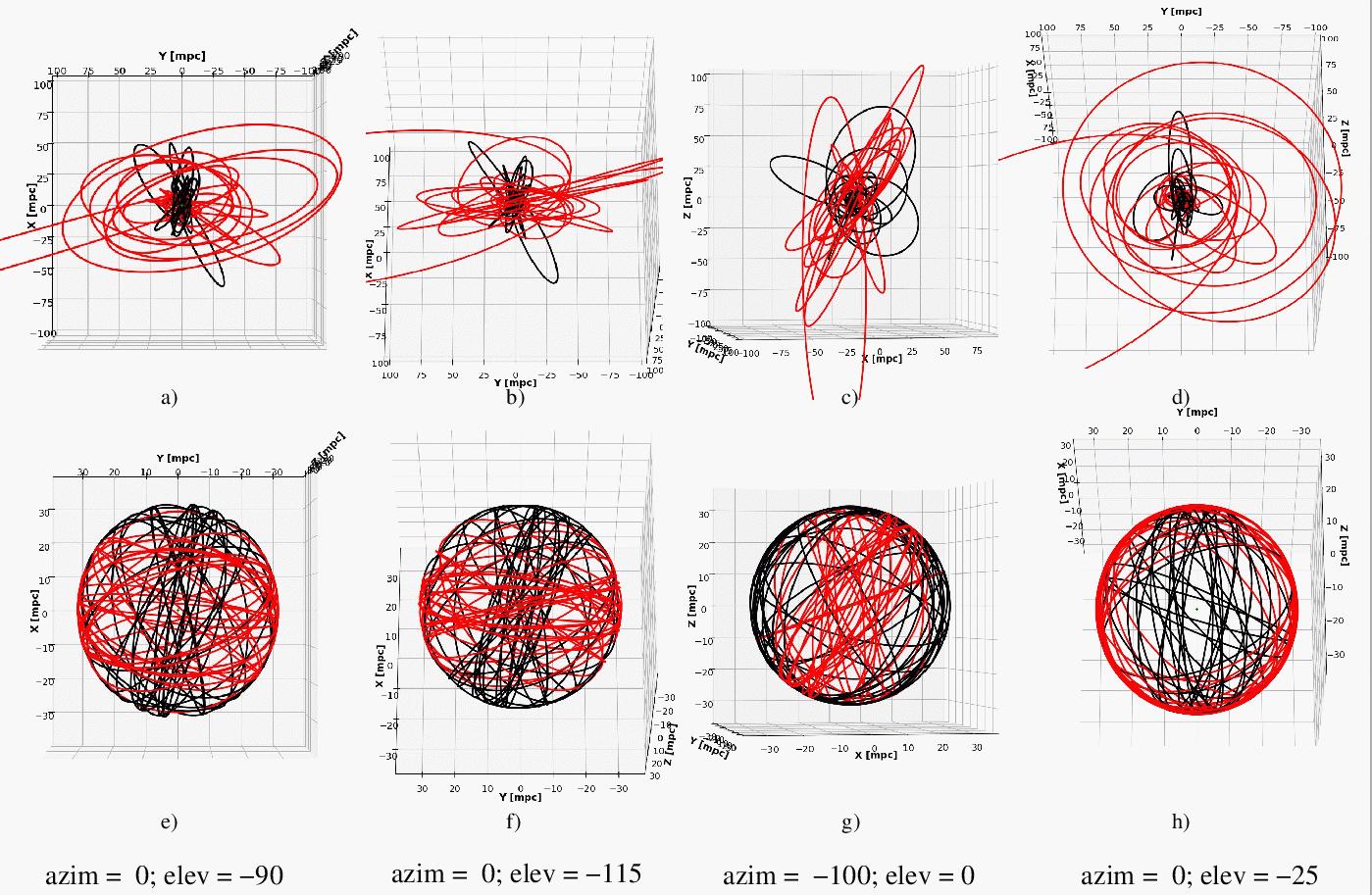}
	\caption{Visualizations of the distribution of all 39 orbits of the S-cluster stars.
In the top row the orbital elements as derived from the observational data
are used. In the bottom row the ellipticities have been set to zero and the
semi-major axes have been set to a constant value. Hence, only the orientation 
angles of the orbits are relevant for the visualization.  
The azimuthal and elevation angles for the corresponding projecting are given.
Panels {\bf a) and e)} show the line of sight view as observed.
Panel {\bf b) and f)} show both disk systems seen edge on.
In {\bf c) and g)} the orbits of the black system are face on, 
and those of the red system are edge on.
Finally, panels {\bf d) and h)} show the red system face on, 
the black system edge on.
We refer also to the supplemented enhanced graphics that
show the projected orbital arrangements in motion.
}
\label{fig:orbits}
\end{figure*}

\begin{figure}
	\centering
	\includegraphics[width=0.7\columnwidth]{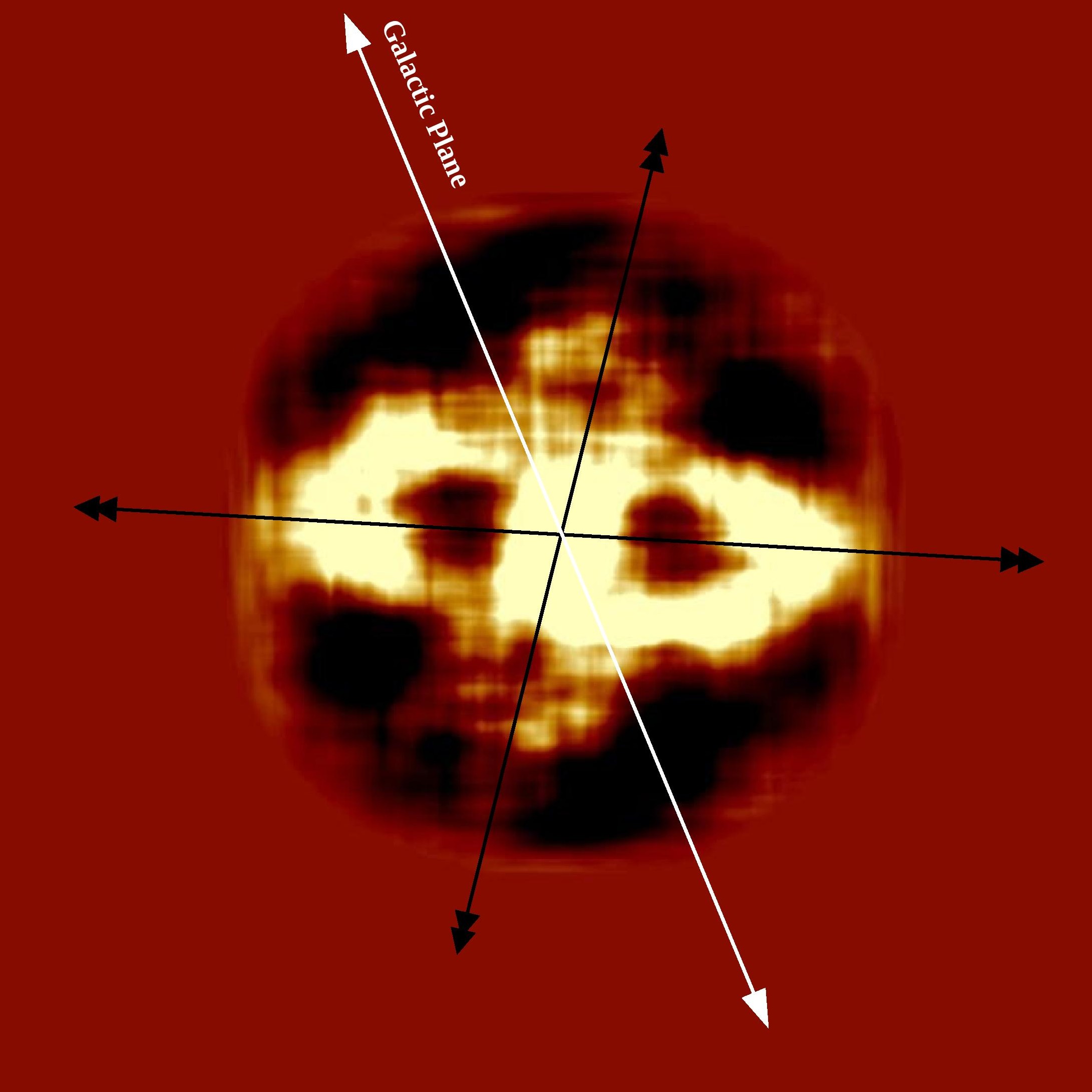}
	\caption{
Smoothed representation of the pole vision image of the circularized orbit distribution 
shown in Fig.\ref{fig:orbits}~f.
We subtracted the distribution expected from 39 randomly oriented orbits.
Here the 39 orbits are generated, assuming a $sin(i)$ uniform distribution for 
the inclinations, and a circular uniform distribution of the longitude 
of ascending node after setting $e = 0$ and $a = const$.
As in Fig.\ref{fig:orbits}~f
the black lines indicate the directions of the disks and the white line
the direction of the Galactic plane.
}
	\label{fig:xshape2}
\end{figure}

In Fig.\ref{fig:orbits} we show the three-dimensional distribution of the
orbits.
In all projections the two organization of two orthogonal disk (black and red)
of the stars is apparent.
The coloring is based on visual inspection of perpendicularity in three dimensions.
In sub-figures a) to d) we show the orbits using the complete set of orbital 
elements. 
In sub-figures e) to h) we show the circularized orbits after the eccentricity 
had been set to zero and the long orbital axes had been set to a 
constant value. In this version, only the orbital angles are preserved
and the bunching into orbital families becomes most apparent.
In sub-figures a) and e) the face on view as seen from Earth is presented.
In this case the black orbital family is seen almost edge on. 
In sub-figures b) and f) the set of orbits has been rotated by 25$^o$
from elevation -90$^o$ to -115$^o$. Here the two orbital families are both seen edge on.
In Fig.\ref{fig:xshape2} we show a smoothed version of the pole vision for the 
circularized orbits in subimage f).
Here the X-shape structure of the two disks can be seen more clearly.
In sub-figures c) and g) the set of orbits has been rotated by -100$^o$ in azimuth (keeping the elevation at 0).
In this case the black orbital family is seen face on while the
red orbital family is edge on.
In figure d) and h) we rotated to elevation -25$^o$ (and azimuth at 0$^o$)
such that the red system is then seen face on and the black system is edge on.

\begin{figure}[htbp!]
	\centering
	\includegraphics[width=\columnwidth]{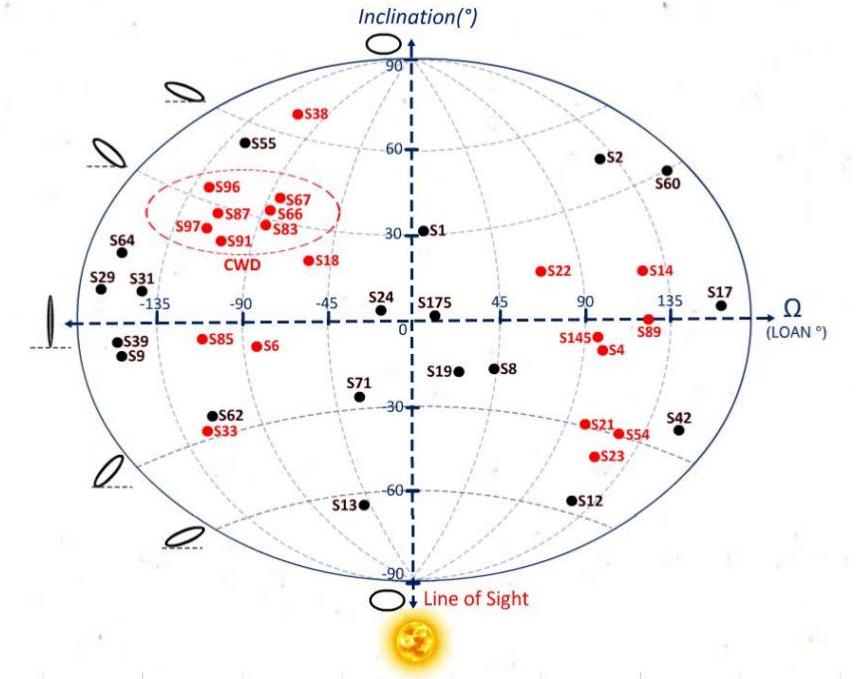}
	\caption{The inclination as a function of the (LOAN)
longitude of the ascending node $\Omega$.
}
\label{fig:orbits3}
\end{figure}

The two orbital disk systems are well separable (see above) but rather thick.
Furthermore the orthogonal X-shaped disk structure becomes apparent only in alternating zones
in the position angle histogram (see Fig.\ref{fig:PAorb})
and in $\Omega$ diagrams as well (see Fig.\ref{fig:orbits3}).
This leads to the fact that they cannot easily be recognized in polar diagrams 
as used by, e.g., \cite{Gillessen2017} in their Fig.12
or \cite{Yelda2014} in their Fig.21.
In Fig.\ref{fig:orbits3} we show the inclination of the stars as a function of the longitude of the ascending
node $\Omega$. The color indicates their membership to either the red of the black disk.
It becomes clear that the two disks cannot easily be identified as the angular momentum 
vectors of the disk members point in opposite directions. 
Comparing Fig.\ref{fig:PAorb} and Fig.\ref{fig:orbits3} 
one can also see that the two disks can better be separated by evaluation the 
in the position angle histogram instead of evaluating the longitude of the ascending node $\Omega$.

However, compared to Fig.\ref{fig:orbits3}, the inclined and face on representations of
the disk members as shown in Fig.\ref{fig:orbits}
are better grouped together, since the direction of the angular momentum 
vector is not relevant in this representation.

Here, looking at circularized orbits as described above
is more successful
in searching for face on orbits that bunch close to the circumference of the 
sky projected distributions like in Fig.\ref{fig:orbits}g or h.

\begin{figure*}[htbp!]
	\centering
	\includegraphics[width=0.9\textwidth]{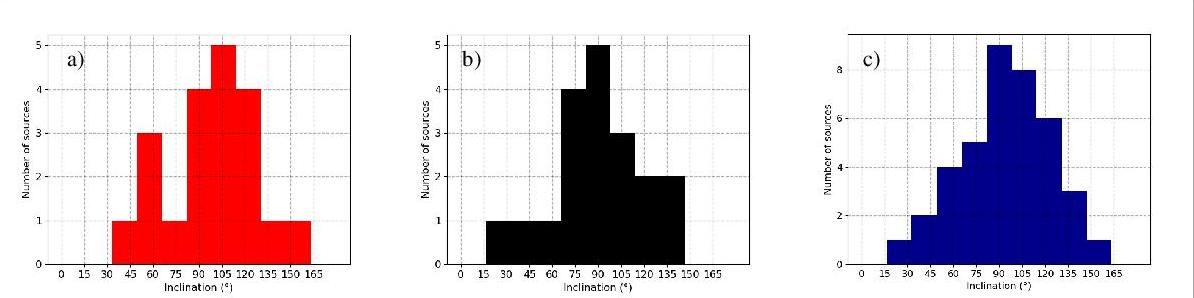}
	\caption{
Inclination angles of all 39 stars with known orbits.
We find that most of the orbits are highly inclined and seen almost edge on.
{\bf a)} Red disk: Inclination angles of all 20 stars, which orbit in the east-west disk.
{\bf b)} Black disk: Inclination angles of all 19 stars, which orbit in the north-south disk.
{\bf c)} All stars in a combined histogram.}
\label{fig:incs}
\end{figure*}

\begin{figure*}[htbp!]
	\centering
	\includegraphics[width=0.9\textwidth]{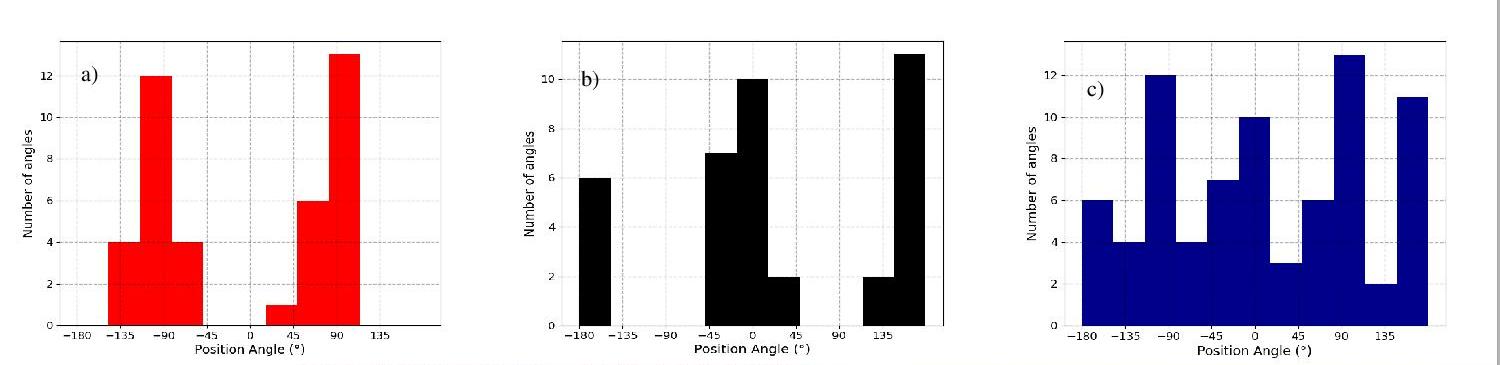}
	\caption{
The distributions of the position angles of the semi-major axes 
of the sky projected orbits shows that the orbits 
of the stars in the red and black system are orthogonal to each other.
{\bf a)}  Position angles of the 19 stars, which orbit in the black north-south disk.
{\bf b)} Position angles of the 20 stars, which orbit in the red east-west disk.
{\bf c)} Position angles of all 39 stars with known orbits.
}
	\label{fig:PAorb}
\end{figure*}

In Fig.\ref{fig:incs}a and b we show the inclinations of the two stellar systems.
On the right hand side of Fig.\ref{fig:incs} we show the distribution of inclinations 
for all stars within the central acrseconds for which we can provide Newtonian orbital fits.
In particular the distributions for the two disks do not follow a $sin(i)$  distribution as one might have expected for a 
fully randomized scenario. 
There is a clear clustering of inclinations around a mean value of $\sim$90$^o$ with 
the bulk of the higher inclined stellar orbits contained within an interval width of
50$^o$ (red disk with a bulk between 80$^o$ to 130$^o$) or even
a width of only 40$^o$ (black disk with a bulk between 70$^o$ to 110$^o$).
In comparison to the expected width of about 100$^o$ for a $sin(i)$ distribution this implies that the two 
separate disk are highly biased towards high inclinations. 
There are also no stars with inclinations in the intervals  0$^o$ to 20$^o$ and 160$^o$ to 180$^o$. 


It follows that the S-cluster stars for which we obtained orbits are organized in
two highly inclined disk systems that are arranged in an X-shape.

\subsubsection{Distribution of orbits in the sky}
\label{section:orbitssky}

Since most stars have high inclinations, the relative orientation of their orbits in the sky can be 
investigated by comparing the position angles of the semi-major axes of their
sky projected orbits.

\begin{figure*}[htbp!]
	\centering
	\includegraphics[width=0.8\textwidth]{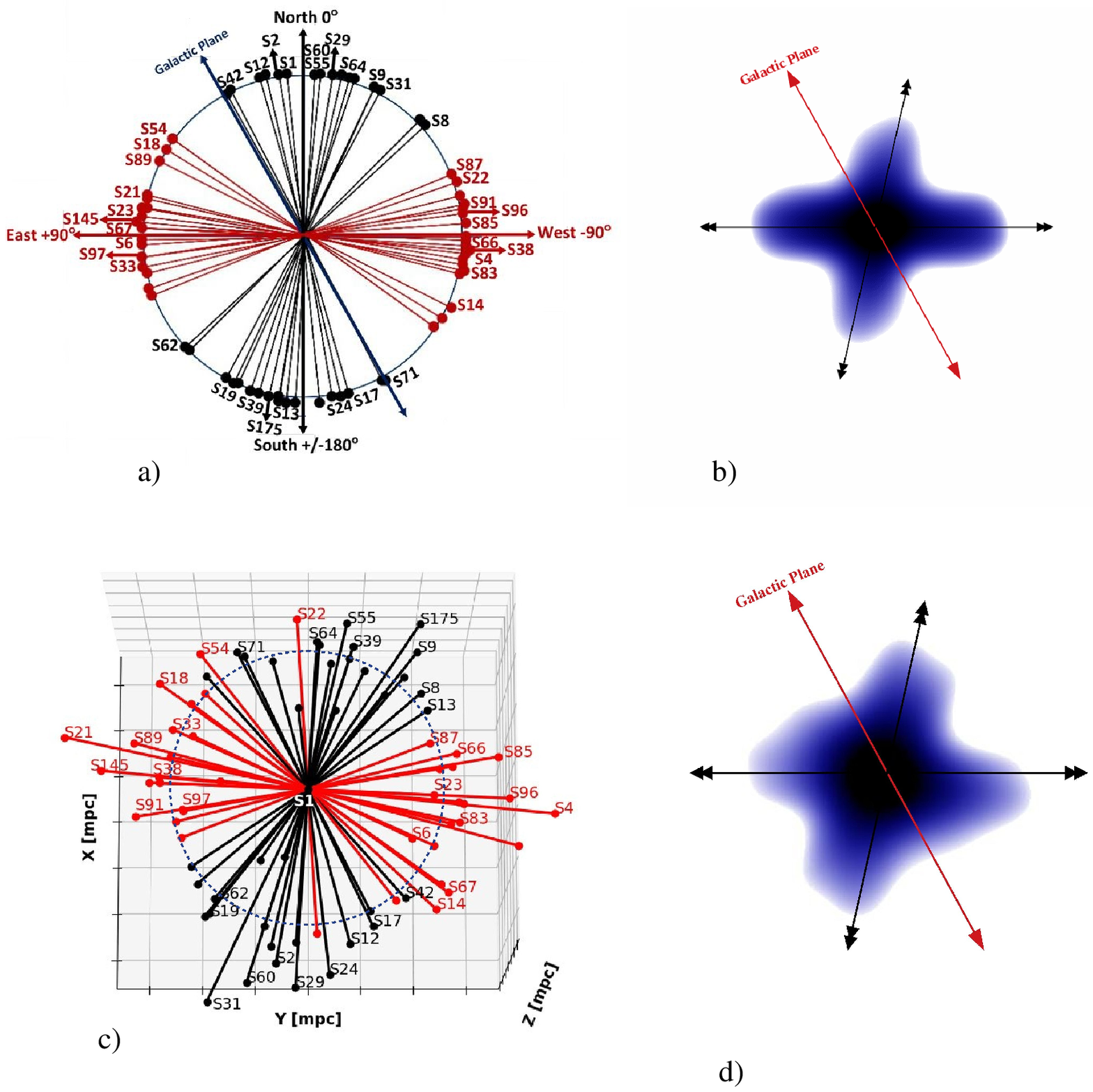}
	\caption{
Angular arrangement of the disks for all 39 stars for which we have orbital solutions.
We labeled the lines for all stars.
{\bf a)}
Here we show for all stars the position angles of the semi-major axes of their
sky projected orbits.
The shape of the two disks is remarkably clear.
{\bf c)}
The position angles of the projected semi-major
axes of the three-dimensional orbits
of all stars after rotation close to the 
pole vision of the system.
The X-shaped is observed in the range between elevation -90$^o$ and -115$^o$ 
but most clearly close to elevation -100$^o$.
The two disks are well separable. 
{\bf b) and d)}
are the same as in a) and c) but smoothed with a circular Gaussian of a width 
corresponding to about 1/5 of the line length in figure section a) and c).
For d) only the position angle line in c) inside the dashed circle have been used for
the convolution.
}
	\label{fig:xshape}
\end{figure*}

In Fig.\ref{fig:xshape}~a) we show the position angles of the
sky projected orbital ellipses in a circular histogram. The orthogonal
red and black orbital families are apparent.
In Fig.\ref{fig:xshape}~b) we show the same diagram consisting of lines indication 
the same position angles but now smoothed with a circular Gaussian with a 
width corresponding to about 1/5 of the line length. Here the representation 
of the line density is enhanced. Both stellar disks have an angle  of about 
45$^o$ with respect to the Galactic plane.

The stars can clearly be separated into two groups (black and red)
that form two stellar disks orientated almost perpendicular to each other.
In Fig.\ref{fig:PAorb} we show how the position angles of the projected orbits are distributed 
for the two disks and for all of the 39 stars. 
Each of the position angles 
is supplemented by a second angle separated by 180$^o$. By this we account for the fact that the stars will 
ascend and descend on their highly elliptical orbits. 
The total number of angles considered in Fig.\ref{fig:PAorb} is 78.

The red disk clusters around $\pm$90$^o$, while
the black disk is concentrated around the angles 0$^o$ and $\pm$180$^o$.
In order to investigate the statistical significance of this arrangements we need to
apply methods that have been developed for directional statistical analysis.
Starting with the multi-modal distribution of position angles, we can apply Rao's spacing test
\citep{Rao2001}.
The test is based on the idea that if the underlying distribution is uniform, then the 
observation of $N$ successive directions
should be approximately evenly spaced. They should show an angular separation of about $360^o/N$.
Large deviations from this distribution, resulting from unusually large spaces or unusually 
short spaces between the observed directions, are evidence for directionality.
The test is more powerful than the Rayleigh test \citep{Durand1958} when it comes to multi-modal distributions.
After placing all 78 position angles on a circle, we performed the test and the resulting 
p-value is 0.01 with a test statistic of 154.12 and a critical statistic of 152.46,
allowing us rejected the hypothesis that the distribution is uniform.
In addition we performed the Hodges-Ajne test \citep{Ajne1968, Bhattacharayya1969},
for uniformity of a circular distribution.
The test is based on the idea that if the number of points in an arc exceeds 
the expected number for a uniform distribution then the hypothesis is rejected.
The implemented Hodges-Ajne test in Python returns either 1 or 0 as a p-value.
Applying it to our position angle distribution we obtained a p-value of 0. 

Hence, they can be separate very well. Thanks to the high inclination of the orbits 
the pole of this distribution (i.e. the region where most orbits cross 
each other) is close to the line of sight and the two stellar systems can be
separated even in their direct projected appearance in the sky.
The same can be done with the projection of the semi-major axes of the 
three-dimensional orbits, as shown in Fig.~\ref{fig:xshape}c.
To get a clear view we rotate the orbital arrangement close to the pole vision 
and use the projected semi-major axes of the orbits.
The smoothed version of the distribution is shown in Fig.~\ref{fig:xshape}d.

We note that the trend for having two orthogonal disks is probably
also continued towards stars with separations to SgrA* smaller than those
of the star S2. The recently found high velocity star S62
\citep{2020Peissker} lies to within 30$^o$ close to the black disk.
This star has separation to SgrA* ranging between 17.8~AU to 740~AU,
compared to S2 with 120~AU to 970~AU.
We expect to find stars close to the red disk with similarly short periods and
small distances to SgrA*.

This shows that the highly inclined stellar orbits can indeed be separated
into two groups that represent two orthogonal disks.

\begin{figure*}[htbp!]
	\centering
	\includegraphics[width=0.9\textwidth]{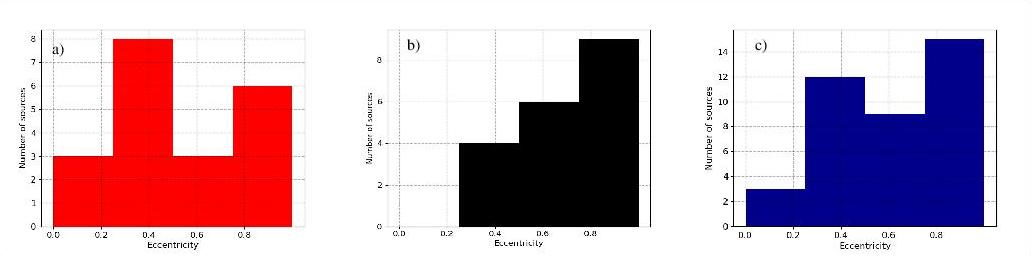}
	\caption{
{\bf a)} 
Eccentricities of the 20 stars, which orbit in the red east-west disk.
{\bf b)} 
Eccentricities of the 19 stars, which orbit in the black north-south disk.
{\bf c)} 
Eccentricities of all 39 stars with known orbits (including the 7 ex-members).
}
	\label{fig:ecc}
\end{figure*}

\subsubsection{Orbital eccentricities}
\cite{Gillessen2017} find that the distribution of eccentricities of the 
S-star cluster is thermal
which is in agreement with Fig.~\ref{fig:ecc}c.
However, this obviously does not necessarily imply that 
the orbits are randomly oriented. 
The two disk systems show that the S-star cluster is highly organized.
In Fig.~\ref{fig:ecc}a and b we show the histogram of eccentricities 
for the two disk systems.
There are only about half the number of sources in the individual histograms,
however, within the uncertainties we find
at least for the black disk a distribution that is consistent with a thermal distribution.
The distribution of the red disk is much flatter and is even biased towards the 
low ellipticity side of the diagram, i.e. towards the less thermal side.
This would imply a more thermal, relaxed distribution as expected from the Hills
mechanism \citep{Hills1988} for the black disk. For the red disk the implication is that it is 
more influenced by a disk-migration scenario as it approaches the less-than-thermal side
of the graph.
This is consistent with the fact the the black disk is more compact - it is confined to within 
a radius of about one arcseconds - and the red disk is significantly larger, 
with stars confined to within 2.3 arcsecond radius.
However, while at first glance, it may be coupled to the clockwise rotating stellar disk 
(CWD) of He-stars \citep{Levin2003, Paumard2006},
it is likely to have a different origin or history 
since the angular momentum vectors for individual stars in each disk point in opposite
directions.
Here, scattering of resonance mechanisms may be more important than at larger 
distances from SgrA* (see discussion in section~\ref{subsub:relax}).

\begin{figure}
	\centering
	\includegraphics[width=\columnwidth]{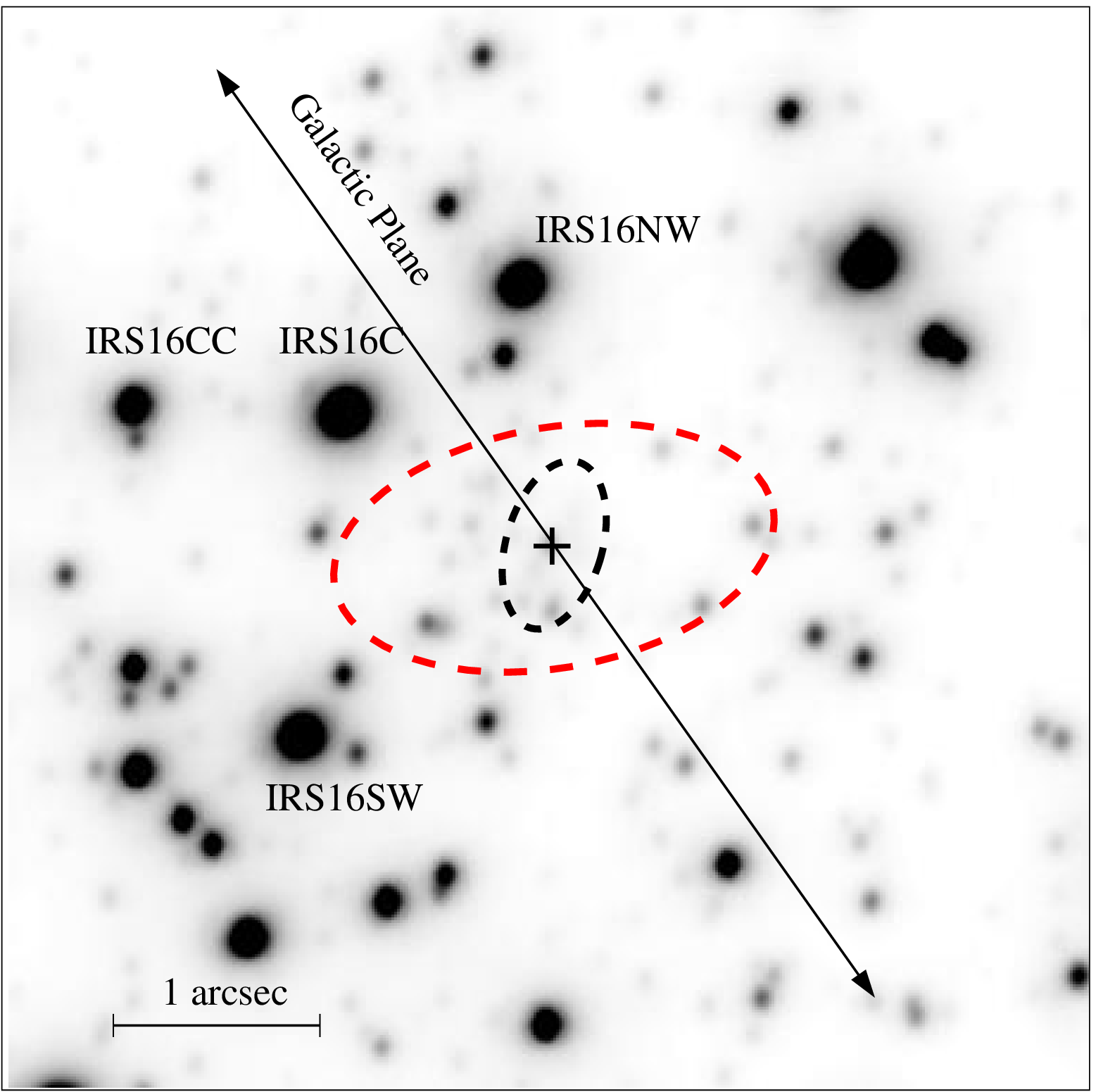}
	\caption{
Location and extent of the red and black stellar disks with 
respect to the Galactic Center stellar cluster and the Galactic plane.
SgrA* is located at the center of the open cross close to S2.
East is to the left, north is up.
The semi-major axis of the black (red) dashed ellipse is about twice median 
of 0.4" or 16~mpc (1.18" or 47~mpc) 
the semi-major axes of all orbits attributed to the black (red) disk system.
The minor axes of the ellipses have been chosen such that they
include the central half of the corresponding disk system orbits.
The red dashed line comprises the bulk of the S-cluster stars.
The epoch of the underlying image is early 2018.
	\label{fig:map2}
}
\end{figure}

It is not clear if and how the more compact black disk is
coupled to the counter-clockwise disk claimed to be perpendicular to the CWD 
\citep{Paumard2006}.
How the two stellar disk are arranged in projection against the sky and the
Galactic Center stellar cluster is shown in Fig.~\ref{fig:map2}.

\subsection{Stellar dynamical considerations}
\label{subsection:dynamics}

 Stars bound to a super-massive black hole interact gravitationally. 
The reason for the non-isotropic distribution of S cluster members 
may be inferred by comparing the characteristic time-scales of 
different dynamical processes (non-resonant two-body relaxation, resonant relaxation) 
with the estimated age of S-stars. For S2 star, \citet{Habibi2017} derive an age of $6.6^{+3.4}_{-4.7}$ based on twelve years of spectroscopic monitoring, with the cumulative signal-to-noise ratio of $S/N>200$, with an upper limit on the formation time of S-stars of $<15\,{\rm Myr}$. This is consistent within uncertainties with the formation time of the clock-wise (CW) disk of young, massive OB/WR stars, $5\pm 1\,{\rm Myr}$, which occupies the region beyond the S cluster at the deprojected distance between $\sim 0.04\,{\rm pc}$ and $0.5\,{\rm pc}$ \citep{genzel2010}. This suggests a common origin of massive OB stars in the CW disk and those of lighter S-stars of spectral type B.

 Recently, a group of NIR-excess compact sources was identified \citep{Eckart2013}, whose spectral properties, in particular for the intensively monitored DSO/G2 object \citep{Gillessen2012,Witzel2014,Valencia2015}, suggest that these could be pre-main-sequence stars of Class I source with an even younger age of $\sim 0.1$--$1\,{\rm Myr}$ \citep{2017A&A...602A.121Z}.
If DSO/G2, G1 object and other NIR-excess sources are pre-main-sequence stars of class I (with the age of $\sim 0.1$--$1\,{\rm Myr}$), then their orbits should also keep dynamical imprints of the initial formation process, e.g. most likely an in-fall of the molecular clump and a subsequent \textit{in situ} star formation \citep{Jalali2014}. In that case, NIR-excess sources could form a dynamically
related group of objects, e.g. their inclinations would be comparable, which can be tested observationally in the future when orbital elements for more objects will be inferred.
In case an additional gas in-fall occurred after the stellar disk formation, its effect 'superimposed' on 
the dynamical effect any  residual disk gas could have had. The evidence for the in-spiral of  
fresh gas is supported by 
\cite{YusefZadeh2013,YusefZadeh2017}
who  identified traces (SiO outflows, bipolar outflows) of the recent  star-formation ($10^{4}$ to $10^{5}$ years ago) 
in the inner parsec. In addition,  the discovery of the population of compact NIR-excess sources 
(DSO, G1  etc.) supports the theory of recent and ongoing star-formation and  molecular gas 
replenishment in the inner parsec.

\subsubsection{Basic dynamical timescales}
\label{subsub:relax}

The population of S-stars consisting of 2 disks is not relaxed, hence
any current configuration is subject to resonant and non-resonant
relaxation processes in the nuclear star cluster. The configuration
of 2 perpendicular stellar disks can be stable over a  timescale
of 10$^8$ years as demonstrated in the simulations by \cite{Masto2019}
that we refer to later. In the current section we mention key
dynamical processes that might have contributed to the
X-structure and so far could have influenced it.
In particular, the resonant relaxation process can lead to the
spread in orbital inclinations in each disk.
An important quantity to understand the dynamics of a stellar system is
the relaxation time scale within which a system reaches a 
statistical equilibrium through stellar interactions.
Persistent torques
acting between the orbits of the S-stars will lead to the rapid resonant relaxation 
of the orbital orientation vectors (“vector” resonant relaxation) and the 
slower relaxation of the eccentricities (“scalar” resonant relaxation).
These mechanisms both act at rates much faster than two-body or
non-resonant relaxation. 
Possible physical sources of orbit perturbations are discussed 
in section \ref{subsub:pertirb}.

\begin{center}
{\it a) Resonant relaxation time-scales}
\end{center}

For calculating typical time-scales, we adopt the relations presented by \citet{HA2006}. 
The {\it non-resonant relaxation} time-scale dominated by two-body  interactions can be expressed as follows,

\begin{equation}
   T_{\rm NR}=A_{\Lambda}\left(\frac{M_{\bullet}}{M_{\star}} \right)^2 \frac{P(a)}{N(<a)}\,
   \label{eq_nonresonant}
\end{equation}  
where $P(a)=2\pi[a^3/(GM_{\bullet})]^{1/2}$ is the Keplerian orbital period and $A_{\Lambda}$ is a dimensionless factor that contains the Coulomb logarithm. $N(<a)$ is the number of stars with the semi-major axes smaller than a given semi-major axis $a$. For the stellar mass $M_{\star}$, we take $M_{\star}=10\,M_{\odot}$, which is the order of magnitude estimated for several S-stars \citep{genzel2010,2017ApJ...847..120H}.

For {\it scalar resonant relaxation}, which changes the value of the angular momentum $|\mathbf{J}|$, we consider the typical time-scale in the following form,

\begin{equation}
   T_{\rm RR,s}=\frac{A_{\rm RR,s}}{N(<a)}\left(\frac{M_{\bullet}}{M_{\star}} \right)^2 P^2(a) |1/t_{\rm M}-1/t_{\rm GR}|\,,
   \label{eq_scalar_resonant_relaxation}
\end{equation}
where the factor $A_{\rm RR,s}=3.56$ is inferred from N-body simulations of \citet{RT96}. The time-scales $t_{\rm M}$ and $t_{\rm GR}$ correspond to the mass-precession and to the general relativity (GR) time-scale, respectively. The mass precession takes place due to the potential of an extended stellar cluster and may be expressed as,

\begin{equation}
  t_{\rm M}=A_{\rm M} \frac{M_{\bullet}}{N(<a) M_{\star}} P(a)\,,
  \label{eq_mass_precession}
\end{equation} 
where the factor $A_{\rm M}$ is of the order of unity. Closer to the black hole associated with Sgr~A*, the GR precession is the dominant effect, which takes place on the time-scale of $t_{\rm GR}$,

\begin{equation}
  t_{\rm GR}=\frac{8}{3}\left(\frac{J}{J_{\rm LSO}} \right)^2 P(a)\,,
  \label{eq_gr_precession}
\end{equation}
where $J_{\rm LSO}\equiv 4GM_{\bullet}/c$ is the angular momentum of the last stable orbit. 

{\it Vector resonant relaxation} keeps the magnitude but changes the direction of the angular momentum $\mathbf{J}$. The time-scale of the vector resonant relaxation can be estimated as,

\begin{equation}
  T_{\rm RR,v}\simeq 2 A_{\rm RR,v} \left(\frac{M_{\bullet}}{M_{\star}} \right) \frac{P(a)}{N^{1/2}(<a)}\,,
  \label{eq_vector_resonant_relaxation}
\end{equation}
where the factor $A_{\rm RR,v}=0.31$ \citep{RT96}.

Another process that induces the eccentricity-inclination 
oscillations is the {\it Kozai-Lidov mechanism}, which involves three 
bodies, i.e. inner binary system (black hole-star) perturbed 
by a stellar or a gaseous disk \citep{Subr2005} or an inner binary 
(star-star) perturbed by the black hole \citep{2016MNRAS.460.3494S}. 
The time-scale of Kozai-Lidov oscillations induced by a self-gravitating disk having the mass 
of $M_{r}$ at the distance of $r$ from the Galactic center is \citep{Subr2005,HA2006},

\begin{equation}
  T_{\rm KL}=2\pi  \left(\frac{M_{\bullet}}{M_{r}} \right)\left(\frac{r}{a} \right)^3 P(a)\,.
  \label{eq_kozai_timescale} 
\end{equation}

For quantitative estimates, we used a specific mass density profile of stars $\rho_{\star}(r)$ to calculate 
time-scales expressed by Eqs.~\ref{eq_nonresonant}--\ref{eq_kozai_timescale}. 
We adopted a broken power-law profile according to \citep{schoedel2009,Antonini2012},

\begin{equation}
   \rho_{\star}=\rho_{0} \left(\frac{r}{r_{\rm b}}\right)^{-\gamma_{\rm s}} \left[1+\left(\frac{r}{r_{\rm b}}\right)^2\right]^{(\gamma_{\rm s}-1.8)/2}\,,
   \label{eq_stellar_density}
\end{equation}
where $\gamma_{\rm s}$ is the inner slope, $r_{\rm b}$ is the break radius, for which we take $r_{\rm b}=0.5\,\rm{pc}$. Setting $\rho_0=5.2 \times 10^5\,M_{\odot}{\rm pc^{-3}}$ gives the integrated, extended mass in accordance with \citet{schoedel2009}, within their inferred range of $\sim 0.5$--$1.5 \times 10^6\,M_{\odot}$ (with the black hole mass subtracted). We consider two cases for the inner slope:
\begin{itemize}
  \item $\gamma_{\rm s}=1.0$, which is consistent with the volume density of the S cluster, $\rho_{\rm S}\propto r^{-1.1\pm 0.3}$ based on the orbits of 15 stars with the semi-major axis of $a\lesssim 0.5''$ \citep{genzel2010},
  \item $\gamma_{\rm s}=0.5$, which represents the overall observed stellar 
distribution in the central parsec \citep{Buchholz2009}.
\end{itemize} 

\begin{figure*}[tbh]
  \centering
     \includegraphics[width=\textwidth]{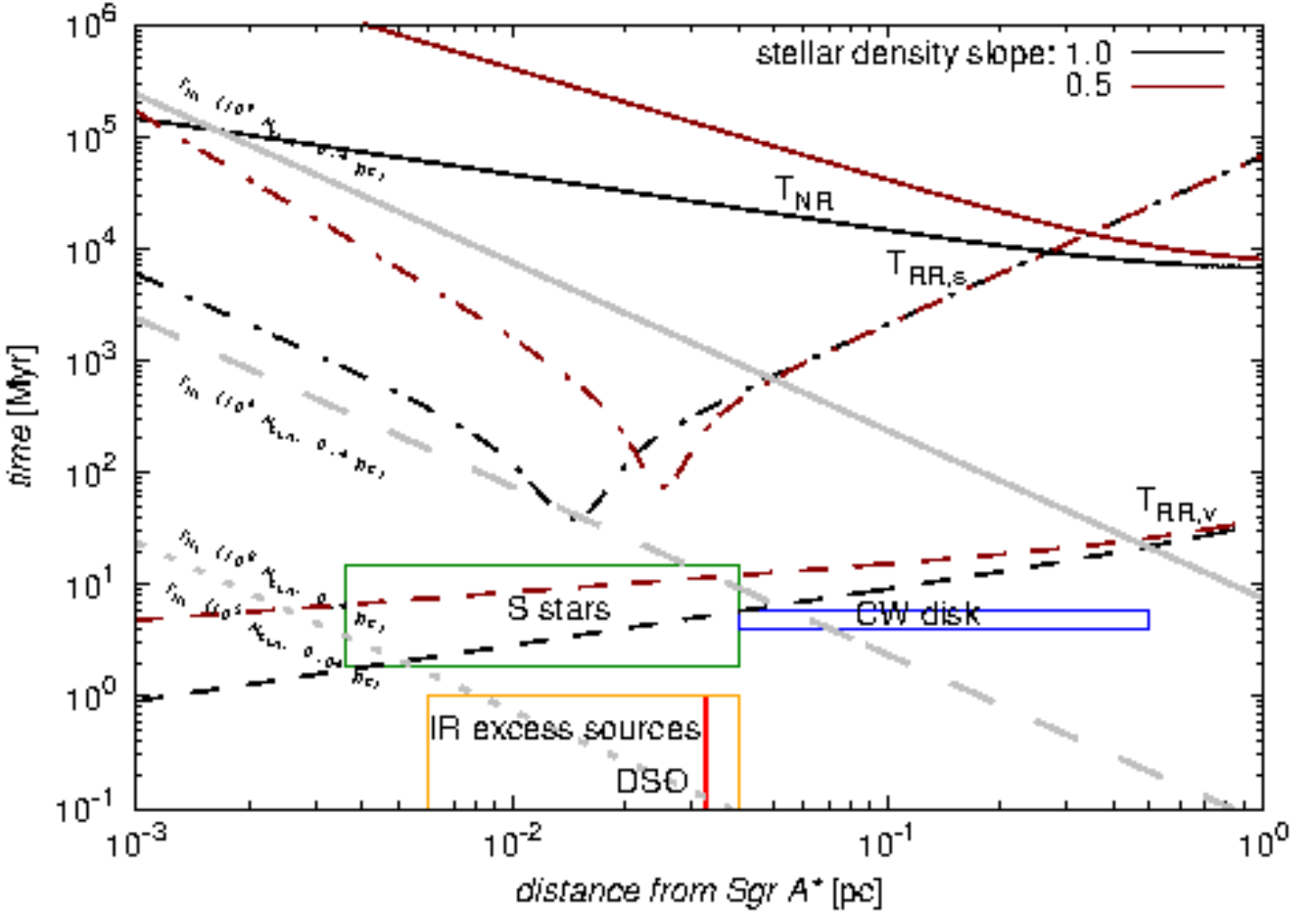}
  \caption{
The time in millions of years (Myr) versus the semi-major axis in parsecs (pc) for the inner slope of the stellar
density distribution equal to $\gamma_{\rm s} = 0.5$ and $\gamma_{\rm s} = 1.0$, which are depicted by 
different colours, dark-red and black, respectively. Different lines correspond to the estimates of 
typical time-scales of dynamical processes operating in the central parsec: $T_{\rm NR}$ corresponds 
to the non-resonant relaxation (solid black and dark-red lines), $T_{\rm RR,s}$ stands for a scalar 
resonant relaxation (dot-dashed black and dark-line lines) , $T_{\rm RR,v}$ for a vector resonant 
relaxation (dashed black and dark-red lines), and $T_{\rm KL}$ for Kozai-Lidov
time-scales (gray solid, long-dashed, and short-dashed lines, depending on the mass and the distance 
of the stellar disk). All the timescales are calculated according to relations 
Eqs.~\ref{eq_nonresonant}--\ref{eq_kozai_timescale}.  for the individual stellar mass of
$M_{\star} = 10 M_{\odot}$ when relevant. The distinct minimum time for scalar resonant relaxation 
corresponds to the semi-major axis, where the GR precession takes over the extended Newtonian-mass precession. 
The values in parentheses next to the Kozai time-scale, e.g. $T_{\rm KL}\,(10^{4}\, M_{\sun} , 
0.4 {\rm pc})$ represent the parameters of the massive gaseous or stellar disk, 
in particular $M_{\rm r} = 10^{4} M_{\odot}$ that is at the distance of $r = 0.4\, {\rm pc}$, see also 
Eq.~\ref{eq_kozai_timescale}. 
The rectangles stand for the distance as well as the determined age of different stellar populations, 
namely S-stars, CW disk stars, and NIR-excess sources, specifically the DSO is represented by the
thick red solid line. The inner radius of the S-cluster box is now represented by the S62 semi-major 
axis \citep{2020Peissker}.
}

  \label{fig_timescales-1}
\end{figure*}

Using the two stellar distributions, we show the time-scale estimates alongside the characteristic stellar structures (S-stars, CW disk, DSO and other NIR-excess sources) in the time--semi-major axis plot, 
see Fig.~\ref{fig_timescales-1}.

We adopt the age constraints of S-stars from the recent spectroscopic
study of Habibi et al. 2017, where they show that S-stars are young
and comparable in terms of age to OB stars from the CW disk.
However, it is premature to claim that these two populations are
identical in terms of age. They could have formed in two separate
star-formation events, with a different dynamical configuration.

The findings in this work, in particular the comparisons in Fig.~\ref{fig_timescales-1}, 
suggest that S cluster has not yet completely relaxed in either resonant or non-resonant way. Short-period S cluster members could have been influenced by vector resonant relaxation, which changed their orbital inclination, especially for more peaked stellar-density distributions with $\gamma_{\rm s} \sim 1.0$, see Fig.~\ref{fig_timescales-1}. However, because of the young age of S-stars comparable to CW-disk stars, vector resonant relaxation is not expected to lead to complete randomization of stellar inclinations for S-stars with larger semi-major axes (longer periods) as has been previously argued to explain the apparent nearly isotropic S-cluster distribution \citep{genzel2010}, which is not confirmed in this work. Scalar resonant relaxation, which influences orbital eccentricities and semi-major axes of stars, takes place on time-scales at least one order of magnitude longer than the age of S-stars. 

Hence, the S-cluster can in principle keep the non-isotropic structure, consisting of two inclined disks embedded within the outer CW disk. This may be hypothesized to originate in the way the S-cluster formed. In particular, the S-stars were likely formed within the infalling cloud/streamer that formed the disk around Sgr~A* upon its impact, as seems to be the case for OB stars that are a part of the CW disk further out \citep{Levin2003}. Due to its age of several million years, the S-cluster is expected to keep the imprints of the original formation mechanism within the gas/stellar disk, which potentially consisted of more inclined streamers. The coexistence of more inclined gaseous disks is also predicted by hydro-dynamical simulations of the star-formation in the Galactic center within an infalling massive molecular cloud. The multiple inclined disks
 may result from an in-fall of a massive molecular cloud or from a cloud-cloud collision \citep{2009MNRAS.394..191H,2013ApJ...771..119A,2013MNRAS.433..353L}. 

\begin{center}
{\it b) Kozai-Lidov oscillations due to a  massive disk}
\end{center}
\label{subsub:stellar}

In addition, the current S cluster distribution can reflect the perturbation by an outer massive stellar or gas disk in the distance range of $0.04-0.5\,{\rm pc}$ which led to Kozai-Lidov-type resonances, i.e. to the interchange between the orbital eccentricity and inclination because of the conservation of $z$-component of the angular momentum, $L_{\rm z}=\sqrt{(1-e^2)}\cos{i}=\mathrm{const.}$

 The Kozai-Lidov process can be induced by a rather massive gaseous disk present in the past. Concerning the gas disk, in 
Fig.~\ref{fig_timescales-1}
 we can see that this would be the case for a very massive disk of $M_{r}=10^{8}\,M_{\odot}$ positioned at $r=0.4\,\rm{pc}$ (outer boundary of the CW disk). The same Kozai time-scale is obtained for a less massive disk of $M_{r}=10^5\,M_{\odot}$ that is closer, at $r=0.04\,{\rm pc}$ (inner boundary of the CW disk), i.e. one order of magnitude closer to the black hole. Such a scenario with a massive gaseous disk that extended to smaller radii in the past than the current stellar disk was studied by \citet{2014ApJ...786L..14C}. In their study, the Kozai-Lidov resonance induced by the disk could explain the current, thermalized distribution of mostly B-type S-stars as well as the presence of more massive OB stars outside the S cluster. In their Fig. 1, the estimate of the age of the DSO/G2 NIR-excess object is $\sim 10^5$--$10^{5.5}\,{\rm yr}$, being consistent with the pre-main-sequence star as studied by \citet{2017A&A...602A.121Z}.

\begin{center}
\it c) Kozai-Lidov oscillations due  to a massive perturber (IMBH)
\end{center}
\label{subsub:massive}

Alternatively, the Kozai-Lidov oscillation on the timescale of the order of $1\,{\rm Myr}$ can develop due to the presence of a massive body--perturber in the inner parsec, in particular the intermediate-mass black hole (IMBH) of mass $M_{\rm IMBH}$ with the semi-major axis of $a_{\rm IMBH}$ and the eccentricity $e_{\rm IMBH}$. Any S-star then behaves as a test body that orbits Sgr~A* and is perturbed by an IMBH further out. The period of the oscillation is \citep{2016ARA&A..54..441N},
 
 \begin{equation}
    T_{\rm KL}^{\rm IMBH}=2\pi \frac{\sqrt{GM_{\bullet}}}{Gm_{\rm IMBH}}\frac{a^3_{\rm IMBH}}{a_{\star}^{3/2}}(1-e_{\rm IMBH}^2)^{3/2} \,.
    \label{eq_KL_IMBH}
 \end{equation}
 
\begin{figure}
     \centering
     \includegraphics[width=7cm]{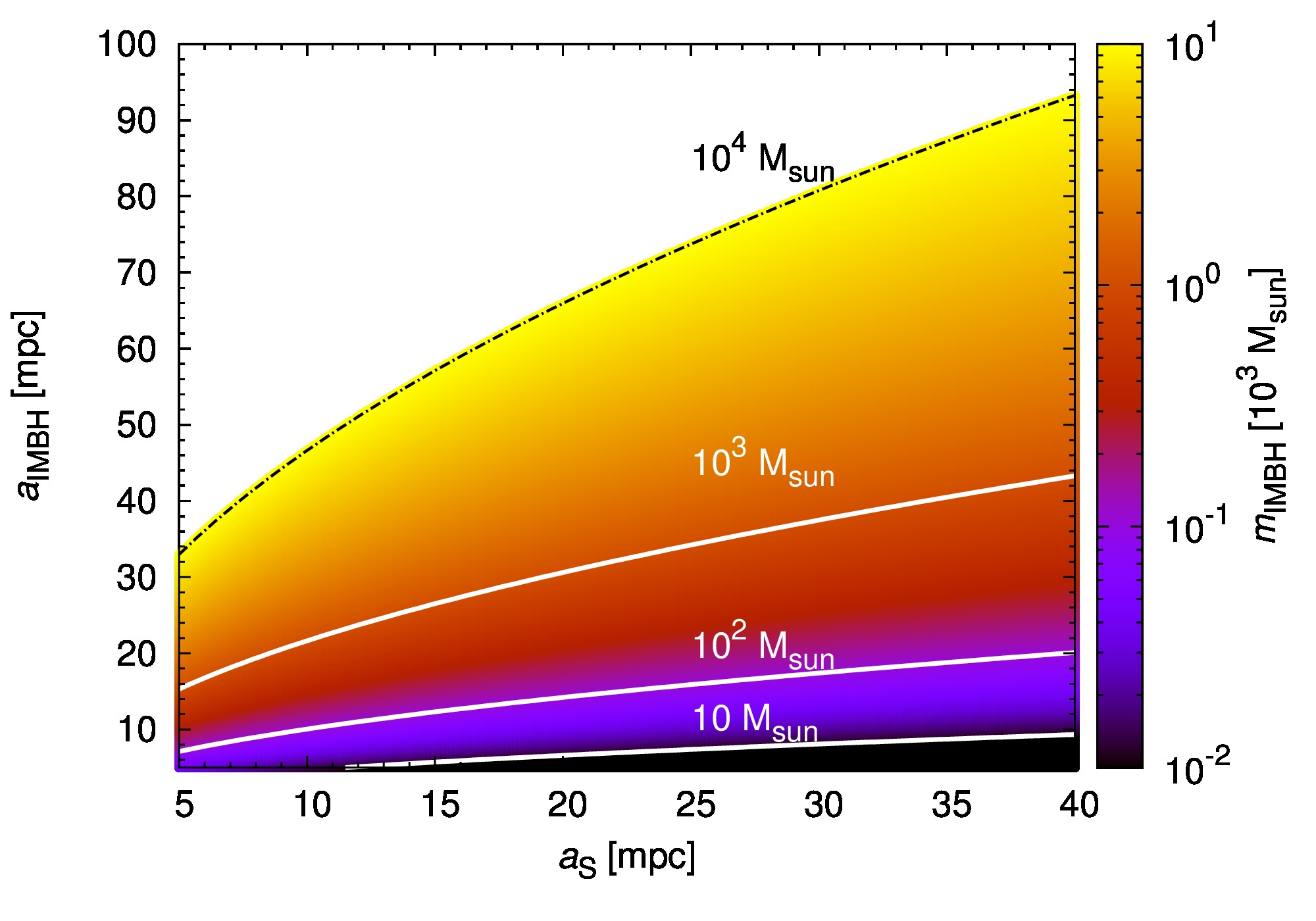}
     \caption{The colour-coded mass of the IMBH as the function of its semi-major axis (circular orbit) 
and the semi-major axis of S-stars for which the timescale of KL oscillations is $1.9\,{\rm Myr}$ 
(lower limit on the age of S-stars). 
}
     \label{fig_KL_IMBH-1}
 \end{figure}

 To get the specific estimates of the mass of the IMBH and its location with respect to the S cluster, we assume the IMBH orbits Sgr~A* on a circular orbit and hence $e_{\rm IMBH}=0$. In Fig.~\ref{fig_KL_IMBH-1}, we show how the location of the IMBH with respect to the S cluster depends on its mass (in the range $10$--$10^4\,M_{\odot}$) in order to induce Kozai-Lidov oscillation in the inclination and the eccentricity during the lifetime of the S cluster. We assumed $T_{\rm KL}=1.9\,{\rm Myr}$. We see that IMBHs with the mass of $m_{\rm IMBH}=10^3\,M_{\odot}$ and lower would essentially have to orbit Sgr~A* within the S cluster on a circular orbit. Only those with $m_{\rm IMBH}=10^4\,M_{\odot}$ and heavier could also be located outside the inner arcsecond to induce the KL resonance in a short enough time for S cluster members.

\begin{figure}
     \centering
     \includegraphics[width=7cm]{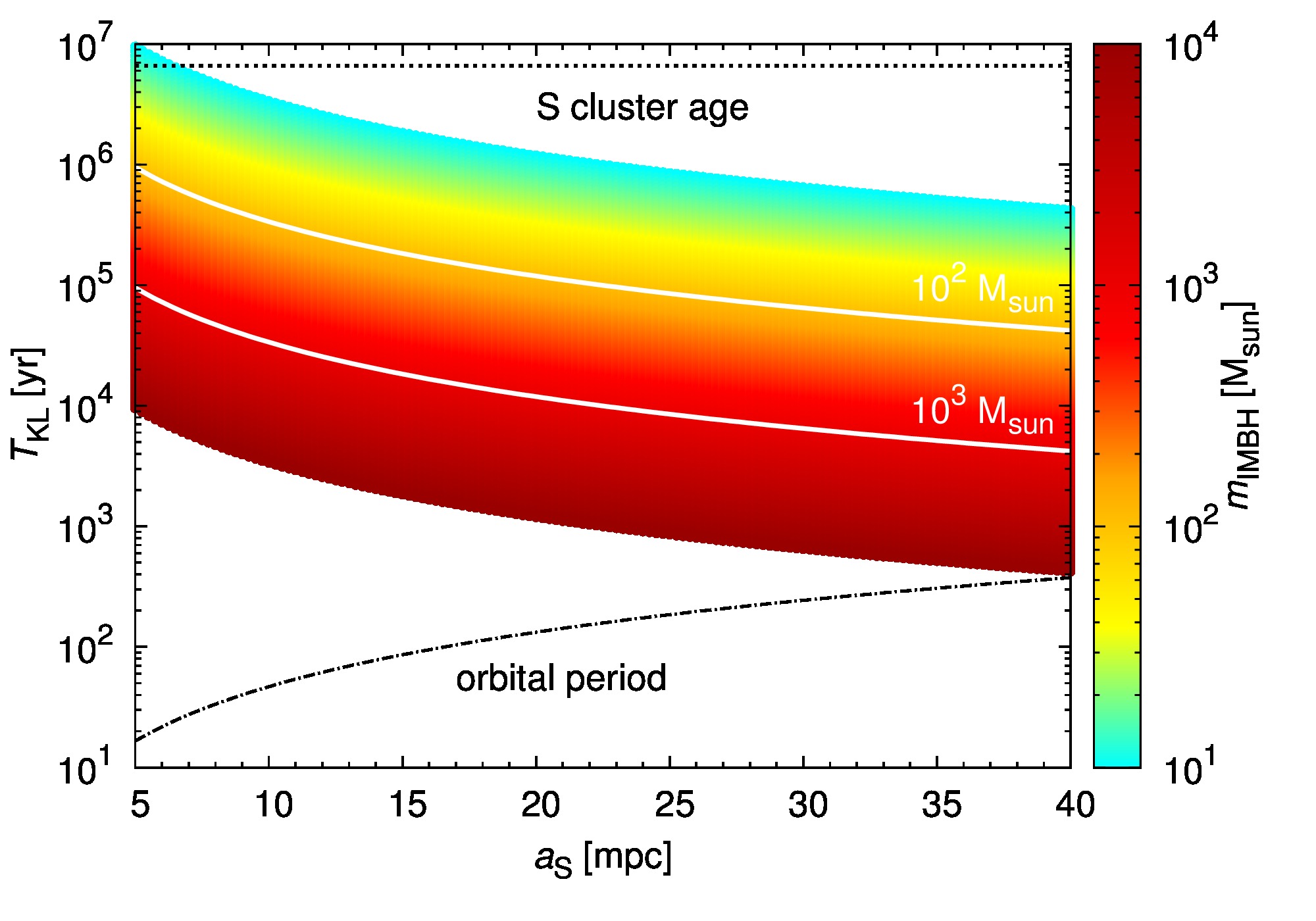}
     \caption{The colour-coded mass of the IMBH as the function of the Kozai-Lidov timescale and of the semi-major axis of S-stars. The IMBH is assumed to have the semi-major axis of $0.04\,{\rm pc}$ and its orbit is highly-eccentric, $e=0.99$.}
     \label{fig_KL_IMBH-2}
 \end{figure}

 From Eq.~\ref{eq_KL_IMBH} it is apparent that the KL timescale can significantly shorten for perturbers-IMBHs that orbit Sgr~A* on a highly eccentric orbit, which can originate due to dynamical scattering in the Nuclear Star Cluster. Specifically, for the IMBH semi-major axis of $a=0.04\,{\rm pc}$ (approximately S cluster length-scale) and the eccentricity of $e_{\rm IMBH}=0.99$, even stellar black holes of mass of the order of $m_{\rm IMBH}=10\,M_{\odot}$ could induce KL oscillation within the S cluster lifetime, see Fig.~\ref{fig_KL_IMBH-2}. For heavier IMBHs, the KL timescale shortens as $T_{\rm KL}\propto m_{\rm IMBH}^{-1}$.

 In conclusion, a massive perturber within or just outside the S cluster can induce the eccentricity-inclination KL oscillations within the S cluster lifetime, i.e. initially disk-like stellar system can get misaligned or initially spherical system can become non-isotropic with respect to the inclination distribution, especially due to KL dependency on initial inclinations -- it applies most significantly to highly inclined stellar orbits with respect to the perturber, $i\sim 40-140$ degrees. Once the system is perturbed due to the KL resonance, it would take at least $T_{\rm RR, v}\approx 10^6$ years for vector resonant relaxation to randomize orbits. Hence, the current S cluster state can reflect a recent perturbation due to the presence of IMBH. 
Although a detailed dynamical modeling is beyond the scope of this paper, the analysis of 
Tisserand's parameter can give a limited insight into the action of a massive perturber near the S-cluster surrounding the SgrA*.
 
\begin{center}
{\it d) Tisserand's parameter}
\end{center}

The Tisserand's parameter is a dynamical quantity that is used to 
describe restricted three-body problems in which the three 
objects all differ greatly in mass.
Tisserand's parameter is calculated from several orbital elements 
(semi-major axis $a$, orbital eccentricity $e$, and inclination $i_b$) 
of a small object and a larger perturbing body, all of which are in 
orbit about a greater central mass. 
This parameter is a dynamical useful quantity 
since it is approximately conserved during an encounter of the two smaller 
bodies.
It therefore allows us to connect the post-encounter dynamical 
properties with the pre-encounter properties 
\citep{Merritt2013}.
In the following we see that the analysis of Tisserand's parameter 
for the S-cluster stars 
suggests that two perpendicular disks can be supported
by a heavy mass just outside the S-cluster, influencing its dynamics.

The Tisserand's parameter can be written as:
\begin{equation}
T=\frac{a_{Pert}}{2 a} + \left[\frac{a}{a_{Pert}}(1-e^2)\right]^{1/2}cos(i_b)
\end{equation}

The ratio between the semi-major axes of a massive perturber 
and the stars is $R'=\frac{a_{Pert}}{a}$.
Assuming the stars are in a disk and had semi-major axes $a=a_{Pert}$/R' and an eccentricity $e$ close to zero.
Then $T=R'/2+R'^{-1/2}$. Then $T$ or $\mu=R'+2 R'^{-1/2}$, respectively, describe the
initial setup. For $R'=1$ one finds $T=3/2$, and $\mu=3$.
For the current orbital elements $(a,e,i_b)$ and the current ratio $R=\frac{a_{Pert}}{a}$ 
one can write  Tisserand's relation for each star as

\begin{equation}
R + 2\left[R^{-1}(1-e^2)\right]^{1/2}cos(i_b) \approx \mu~~~.
\end{equation}

Here, $i_b$ is the inclination of the stars with respect to the plane in which the massive perturber orbits the central mass and
$R=\frac{a_{Pert}}{a}$ is the current ratio between the semi-major axes of a
massive perturber and the stars.
This expression can be rewritten as:

\begin{equation}
cos(i_b)=\pm \sqrt{\frac{R(\mu-R)^2}{4(1-e^2)}}
\label{eqcos}
\end{equation}

\begin{figure}
	\centering
	\includegraphics[width=\columnwidth]{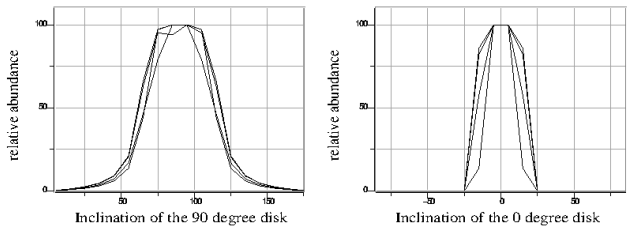}
	\caption{
The peak normalized distributions of inclinations $i_{obse}$ 
that can be derived from
the $i_b = 90^o$ (left) and the $i_b = 0^o$  (right) via 
Tisserand's parameter using the  
observed distribution of eccentricities.
}
	\label{fig:normaldis}
\end{figure}

This relation has simple solutions for cases in which  $\mu \sim R$ with
$\mu$ now containing information on the initial conditions $R'$.
For $R'/R>1.0$ one can reproduce the $i_b=90^o$ disk and
for $R'/R<1.0$ one can reproduce the $i_b=0^o$ disk.
If one allows the current ratio $R$
to vary by a few percent and uses the distribution of observed 
eccentricities as an input, one can reproduce the distribution
of observed stellar inclinations for the $i_b=90^o$ disk.
Compared to the $i_b=90^o$ disk the inclination distribution 
for the $i_b=0^o$ disk turns out to be narrower.
Both distributions are shown in Fig.\ref{fig:normaldis}.
On the sky we observe both disks such that their inclinations 
towards the observer are both close to the observed 
inclination $i_{obs}=90^o$, hence,
the two distributions are superimposed if derived 
from observations of the central arcsecond in total.

\begin{figure}
	\centering
	\includegraphics[width=\columnwidth]{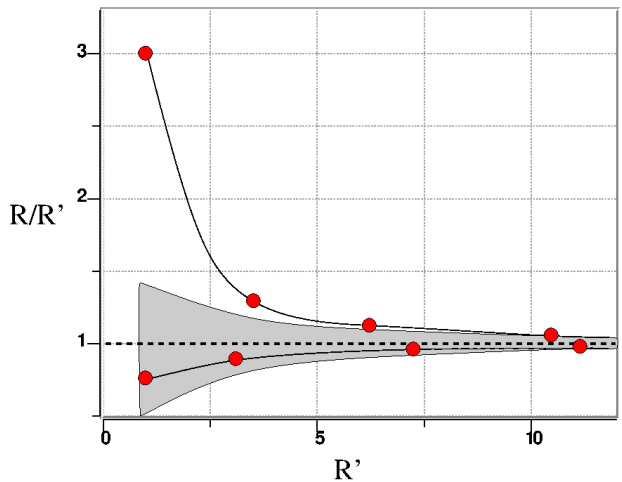}
	\caption{
The current ratio $R$ in relation to the initial ratio
$R/R'$ as a function of the initial ratio $R'$.
The grey areas shows the range covered by 3 times the actual range by
which the current ratio $R$ is allowed to carry in order to explain
the distribution of inclinations.
}
	\label{fig:calcRRs}
\end{figure}

Solving equation \ref{eqcos} for increasing values of $R'$, 
one finds that the value for the current ratio $R$ approximates 
that for initial ratio $R'$.
In Fig.~\ref{fig:calcRRs} we plot the current ratio $R$ in relation to $R'$.
The top graph shows how $R/R'$ approximates unity for the $i_b=90^o$ disk
as listed in Tab.1.
The bottom graph shows the same for the $i_b=0^o$ disk.
For values  $R'\ge6....8$ the difference between the two ratios drops 
so that $R/R'$ gets close to unity  to within less than 
about 3...5 times the width by which we need to let $R$ vary to explain the
observed distributions of inclinations (see Fig.~\ref{fig:calcRRs}).
This means that for these cases the initial conditions are very similar to the 
current conditions and the two orthogonal disks may be populated by objects 
with rather similar dynamical properties.
Hence, we find as a result the analysis of Tisserand's parameter
that two perpendicular disks can be supported
by a heavy mass just outside the S-cluster, influencing its dynamics.
This fits well with IRS13E being a possible disturber of the S-cluster star.
A discussion is given in section \ref{subsub:pertirb}.

Under the influence of a massive perturber the
eccentricity and inclination of the stars may vary periodically 
with the star’s argument of periapsis $\omega$
under conservation of $(1 - e^2 ) cos(i)$. The time scale for these
'Kozai-Lidov cycles' is of the order on 10$^6$ years for the S-stars
within the central 1 to 3 arcseconds if the mass of the perturber 
is of the order of $10^3$ to $10^4$ \solm (see text and equation 8.175 in
\cite{Merritt2013}).
However, there is no specific time scale associated with Tisserand's parameter 
and the formation or conservation of the system.
If at the time of the formation of the stellar disk 
the stars had the observed configuration,
then they will all satisfy 
equation~\ref{eqcos} from the start, and at all later times,
until some other perturbation acts.
Stars that are on orbits that do not satisfy the relation will be
removed or associated by one of the disks a few 
Lidov-Kozai times scales or the resonant relaxation 
time scales (see above).
Two orthogonal disks will be supported by Tisserand's relation and 
the interrelation of stellar angular momenta as described by 
equation~\ref{eqcos}.

\begin{table}[htb]
\caption{Parameters for the disk solutions}
\begin{center}
\begin{tabular}{ccccc}\hline \hline
$i_b$ &$R'$& $R/R'$& $\mu$& $\Delta~R$ \\
\hline
90 & 1   & 3.00& 3.0  & 0.15\\  
90 & 3.5 & 1.30& 4.0  & 0.070 \\
90 & 6.2 & 1.13& 6.6  & 0.042\\
90 & 10.4& 1.05&10.7  & 0.018\\
0  & 1   & 0.77& 3.0  & 0.13\\  
0  & 3.5 & 0.90& 4.0  & 0.070 \\
0  & 7.2 & 0.97& 7.5  & 0.033\\
0  & 11.1& 0.98&11.4  & 0.018\\
\hline \hline
\end{tabular}
\end{center}
Listed are: $i_b$, the inclination of the stellar disks to the orbit of
the perturber; $R'$, initial ratio of the semi-major axes of the stars in the
disk and the perturber; $R/R'$, current ratio  of the stellar disk 
in relation  to the initial ratio;  $\mu$, value of Tisserand's parameter
that is expected to be preserved; $\Delta~R$, half the variation width
of the current ratio of semi-major axes.
\end{table}

\subsubsection{\bf Possible sources of perturbation}
\label{subsub:pertirb}
The strong vertical resonances expressing themselves via the
X-shaped  structure in the stellar distribution can be the 
result of a resonant relaxation process
solely determined by the mean field of stars in the cluster.
Furthermore, the growth of the Galactic bar could trigger Inner 
Lindblad resonances, in which the stars are lifted into 
higher amplitude orbits  (\cite{Quillen2002, binney&tremaine2008}, page 539).
However, one  may identify possible sources of perturbation that 
imposed these resonances or influenced the relaxation process.

{\bf Possibility 1:}
The B-stars of the S-cluster are estimated to have an age less than 15 Myr.
However, star S2 has an age of about 7Myr, which is compatible with the 
age of the clockwise rotating disk of young stars in the GC. 
It is quite likely that S-stars formed almost simultaneously as 
the OB/WR stars that are part of the CW disk \citep{Habibi2017}. 
It is thought that the CW-disk stars formed \textit{in situ} in a 
massive gaseous disk \citep{Levin2003}. 
The origin of this disk could have been a massive molecular cloud 
with the radius of $\sim 15\, {\rm pc}$ and the impact parameter 
of $\sim 26\,{\rm pc}$, which was tidally disrupted, spiraled in 
and subsequently formed an eccentric disk \citep{2012ApJ...749..168M} 
where stars formed. It cannot be excluded that the stellar disk, which 
previously extended below $0.04\,{\rm pc}$ where the S cluster is 
located now, was perturbed by an in-fall of another massive molecular 
cloud that formed a disk or a ring with an inclination of $\beta$ with 
respect to the stellar disk \citep{2013MNRAS.436.3809M,2016ApJ...818...29T}. 
The influx of gas that lead to their formation 
also induced the perturbances in the young S-cluster resulting in vertical 
resonances that relaxed to the structure seen today.
According to the numerical simulations of \citet{2013MNRAS.436.3809M} 
and \citet{2016ApJ...818...29T}, the precession driven by the gas disk 
in the inner $0.5$ pc on the stellar disk can significantly increase the 
stellar inclinations within a few million years, which leads to the disk 
tilting and/or warping. Since the precession is faster for outer disk parts, 
$T_{\rm prec}\propto a_{\star}^{-3/2}$, the S cluster could in principle 
represent a ``primordial'' disk part with two perpendicular streamers that 
have warped to form the CW disk at larger distances. 
Further warping at largest distances led to the disk dismembering which 
can account for $\sim 80\%$ of OB stars that are not a part of any disk 
\citep{2014ApJ...783..131Y}. Two innermost nearly perpendicular stellar 
disks are in a dynamically very stable configuration, since the mutual disk 
precession $T_{\rm DISC}\propto \cos{\beta}^{-1}$ around each other goes 
to infinity for the inclination $\beta$ that approaches $90^{\circ}$.

{\bf Possibility 2:}
Since the orbits are clearly not fully randomized a massive IMBH 
within the S-cluster can probably be excluded (see comments in the 
introduction and see \cite{Gualandris2009}).
However, an IMBH as a massive perturber well outside the S-cluster 
could provide a longterm 
influence  on the orbits resulting in vertical resonances.
The analysis of the Kozai-Lidov oscillations  and Tissarand's parameter 
both suggests that a massive perturber
may have influenced the stellar dynamics in the central arcsecond.
For an initial ratio of the semi-major axes of the stars and the 
perturber of $R'\ge6....8$ the ratio between the initial and the 
current ratio becomes unity.
In this case the dynamical situation may not have changed very much since the 
system was set up.
Assuming that the semi-major axes $a$ of the stars can be taken as a measure of 
the radius of the S-star cluster, i.e. $a=0.5''$ then $R'\sim R \sim 3.5''$
(0.13~pc projected distance).
This fits well with the separation of IRS13E to SgrA* and the S-cluster star.
IRS13E lies to within $\sim$15$^o$ in one of the stellar disks.
If may harbor an up to 10$^4$\solm IMBH and a few hundred solar masses of stars
\citep{Tsuboi2017, Schoedel2005, Maillard2004, krabbe1995}.
The analysis of the Kozai-Lidov oscillations and Tissarand's parameter 
then indicate that the 0$^o$ deg 90$^o$ disk 
in section \ref{subsub:relax}~d)
can to first order be identified as the red and black disk
as discussed in sections \ref{section:orbitssky} and \ref{section:orbitsspace}.

The possibility of different coexisting stellar disks in the Galactic Center
has also been discussed theoretically by \cite{Masto2019}.
Here, the authors simulate multiple stellar disks in the central stellar 
cluster. Each disk is added after  100 Myr. In particular, in the 
bottom panel their Fig.~1, one can see that even after 100~Myr two separate 
stellar disks can still be distinguished. The first one got thicker but is 
still well defined.

\section{\bf Summary and Conclusions}
\label{sec:summary}

We present a detailed analysis of the kinematics of the stars in the innermost stellar cluster
for which we have orbital elements.
The high velocity S-cluster stars
orbit the super massive black hole SgrA* at the center of the Milky Way.
The distribution of inclinations and position angles of the sky projected 
orbits deviate significantly from a uniform distribution which one would have expected 
if the orientation of the orbits are random.
The stars are arranged in two stellar disks that are perpendicular to each other and 
located with a position angle of about  $\pm$45$^o$ with respect to the Galactic plane.
The distribution of eccentricities of the 
inner (black) north- south disk system suggests that it is 
relaxed and thermal as it may be expected from the 
Hills mechanism.
The east-west (red) disk system is 
more influenced by a disk-migration scenario as it approaches the less-than-thermal side
in the distribution of the eccentricities.

While it cannot be excluded that the red disk is the inner 
part of the the clockwise rotating disk (CWD) of He-stars \citep{Levin2003, Paumard2006} or is connected to it,
the black disk system is much more compact and possibly more thermally relaxed.
It is uncertain if or how it is
connected to the large central cluster.

Since the angular momentum vectors of the stars in each disk point in opposite
directions, i.e. the stars in a given disk rotate both ways, it appears 
to be unlikely that the origin or history of these stars is the same as 
the one discussed for the massive young stellar disks containing the He-stars
\citep{Levin2003, Paumard2006, Lu2009, Lu2013, Yelda2014}.
Most likely the S-cluster structure is strongly influenced by
the Kozai-Lidov resonances or vector resonant relaxations.

This prominent X-shaped arrangement is most likely a result 
of the interaction of stars among each other
that can be described via the resonant relaxation process.
An estimate of the resonant relaxation time scale indicates
that the structure started to evolve into the current X-shape 
in the same time interval during which most of the young in the
central stellar cluster were formed.
The presence of a highly ordered kinematic structure at the 
center of the nuclear stellar cluster and in the 
immediate vicinity of the super massive black hole SgrA* also indicates,
that for a very long time no major perturbation of the system occurred that 
could have lead to a randomization of the stellar orbits in the central arcsecond.

\section{\bf Enhanced Graphics}
\label{sec:enhanced}
In all three presentations the super-massive black hole SgrA* is located
at the center of the three-dimensional arrangement.
The animations hold for a short while at the positions under which the
orbital configuration can be seen in special projections:
Red system face on, line of sight view, pole vision, and red system edge on.
Labels in units of milliparsec are given at the edges.
In all animations we list the corresponding azimuth and elevation angle,
with 0$^o$ and -90$^o$, respectively, being the line of sight direction.
In {\it fullorbits.gif} we show the full three-dimensional orbits
using all orbital elements. The dots on the orbits indicate the position of
the star. The corresponding time is given in a text line above.
{\it circularized.gif:} Here, the ellipticity has been set to zero.
{\it normalized.gif:} The ball of wool configuration;
Here, semi-major axes have been set to a constant and
the ellipticity has been set to zero.

\vspace*{0.5cm}
{\bf Acknowledgements:}
We are grateful to Anna Luca H\"ofling, who prepared the enhanced
graphics as an intern in the infrared group at the
I.Physikalisches Institut of the University of Cologne.
We received funding from the European Union Seventh Framework Program (FP7/2007-2013)
under grant agreement No. 312789 - Strong gravity: Probing Strong Gravity by Black
Holes Across the Range of Masses. This work was supported in part by the Deutsche
Forschungsgemeinschaft (DFG) via the Cologne Bonn Graduate School (BCGS), the Max
Planck Society through the International Max Planck Research School (IMPRS) for
Astronomy and Astrophysics, as well as special funds through the University of
Cologne and CRC~956 - Conditions  and Impact of Star Formation under project A2. 
MZ acknowledges the financial support by the National Science Centre,  Poland, 
grant No.~2017/26/A/ST9/00756 (Maestro 9).
We thank the German Academic Exchange Service (DAAD) for support under 
COPRAG2015 (No.57147386) and the
the Czech Science Foundation - DFG collaboration (EC 137/10-1).

\appendix
\section{Field of view effects}
\label{appendixfield}
By restricting the field of view towards a central section 
with radius $\Delta s$$\sim$0.5'' covering the surface $4(\Delta s)^2$ 
in the sky one introduces a bias towards higher inclinations. Sources with low inclinations
with orbits outside the selected area and sources with large three dimensional distances from the
center pass through the selected area only if their orbits have high inclinations.
We assume that the central volume is $16(\Delta s)^3$ and the volume attributed to the 
outer stars within the column $4(\Delta s)^2$ is $8~r~(\Delta~s)^2$.
The central arcsecond is dominated by young stars.
Early type stars are abundant within the central 5 arcsecond radius (i.e. 10$\Delta~s$) of the nuclear cluster
with a surface (volume) density dropping with an exponent -1.8 (-2.8) \citep{Buchholz2009}.
Taking the volume density at 3''-4'' radius, i.e., at 6 to 8 times $\Delta~s$  then the number ratio
of stars between those that are within the volume of the central arcsecond and those that are only in
projection in that region is:
\begin{equation}
(\frac{\Delta s}{(6....8) \Delta s})^{-2.8} \frac{16 (\Delta s)^3}{8 (\Delta s)^2 10 \Delta~s}=(30....67)~~.
\end{equation}
Hence, the bias is only of the order of a few percent and the clustering towards 90$^o$
inclination can be fully attributed to the intrinsic properties of the stellar orbits.

In summary these findings indicate that independent of the 3-dimensional orientation the 
determined stellar orbits in the S-cluster are preferentially seen edge on.

\section{Biases due to incomplete orbital coverage}
\label{appendixbias}

\cite{oNeil2019} discuss the influence of orbital elements resulting from fits to 
data that only cover the orbits in an incomplete way.
They introduce an observable-based prior (OBP) paradigm and the corresponding bias factors 
with respect to uniform priors (UP).
In our case the orbital coverage of the fitted orbits indicates three groups: 

The first group, that contains stars with 40-100\% orbital coverage, has no difference 
between the bias factor of uniform priors UP 
and the bias factor of the OBP for the black hole mass and the distance to the GC. 
The second group, which contains stars with 20-35\% orbital coverage, has a difference 
of 0.3$\sigma$ between the bias factor of UP [0.5-0.8 $\sigma$]
and the bias factor of OBP [0.2 - 0.5 $\sigma$] for the case of the SMBH mass and the distance to it. 
The last group, which contains stars with 5-15\% orbital coverage, has a difference 
of [0.51-0.6 $\sigma$] between the bias factor of UP [0.98-1] and 
the bias factor of OBP [0.4-0.47 $\sigma$] for the case of SMBH mass and the distance to it.
The bias factor difference for the orbital elements \citep{oNeil2019}
was only done for the case of 16\% orbital 
coverage, i.e., valid only for the last group. 
Here, a value of 1 indicates high biased and a value of -1 is low biased (see \citet{oNeil2019}). 


In summary: The inclination bias factor difference is extremely small. For 19 stars only,
 we find a 0.6$\sigma$ difference.
Hence, our finding that the stars are preferentially on highly inclined orbits is unaffected 
by this bias. Also, the orbital elements we derive for the stars are only effected in a 
minimal way.


\bibliography{sclusterrAPJ}


\begin{table}[htb]
\caption{Orbital Elements for Stars in the Black Disk}
\begin{center}
\begin{tabular}{rrrrrrrrrrrrrrrrrrrrrrccccc}
\hline \hline
1. & 2. & 3. & 4. & 5. & 6. &  7. &  8. &  9. &  10. & 11. &  12.&   13. \\ \hline 
Star & a & $\Delta$a & e & $\Delta$e & i & $\Delta$i   &$\omega$ & $\Delta$$\omega$ & $\Omega$ & $\Delta$$\Omega$ & t$_{clos}$ & $\Delta$t$_{clos}$ \\
     &[mpc]& [mpc]     &   &           &[deg]&[deg]&[deg] &[deg]          &[deg]  &[deg]          &   [yr]    &   [yr]   \\
\hline
S1&	22.675&	0.257&	0.665&	0.003&	121.066 & 0.401& 109.893& 0.458&352.484&0.286 & 2000.261& 0.001 \\
S2&	5.034	&0.001	&0.887	&0.002&	137.514	&0.401	&73.416	&0.745	&235.634&1.031& 2002.390& 0.020\\
S8&	16.637&	0.182&	0.768&	0.022&	75.057&	0.573&	337.931& 2.120&	317.075	&0.630&	1979.216&	0.037\\
S9&	11.125&	0.030&	0.791&	0.036&	81.876&	0.458&	137.854&0.573&	158.079	&0.229&	1972.924&	0.023\\
S12&	11.962&	0.105&	0.906&	0.003&	33.060&	0.516&	311.173&0.802	&236.173&1.146&	1995.881&      0.001\\
S13&	9.580&	1.264&	0.415&	0.030&	24.694&	7.219&	256.513&11.459&	47.842&	15.126&	2004.015&	0.507\\
S17&	13.037&	0.794&	0.421&	0.020&	95.799&	0.172&	319.481&3.495&	194.118&1.432&	1991.906&	0.067\\
S19&	11.122&	3.130&	0.626&	0.090&	72.021&	2.807&	131.093&12.261&	337.415&4.469&	2004.275&	0.004\\
S24&	45.115&	7.475&	0.682&	0.061&	95.226&	4.240&	244.596&3.151&	14.381&	1.604&	2023.963&	0.311\\
S29&	34.694&	3.803&	0.335&	0.078&	100.955	&0.688&	331.341&11.975&	171.257&1.432&	2054.568&	4.322\\
S31&	16.582&	4.514&	0.521&	0.151&	108.919&10.256&	321.487	&24.603	&145.990&19.882&2019.201&	1.132\\
S39&	13.919&	2.068&	0.831&	0.042&	86.058&	13.002&	36.784&	9.339&	159.282&0.688&	1999.108&	0.338\\
S42&	38.562&	4.057&	0.644&	0.043&	67.666&	0.802&	37.930&	2.578&	206.379&1.031&	2011.876&	0.716\\
S55&	4.360&	0.002&	0.740&	0.010&	141.692&1.604&	133.499&3.896&	129.890&4.183&	2009.310&	0.030\\
S60&	20.369&	1.799&	0.832&	0.033&	130.806&2.979&	42.743&	11.688&	193.774&17.189&	2021.883&	1.103\\
S62&	3.603&	0.002&	0.980&	0.000&	61.765&	0.057&	45.034&	0.057&	112.414&0.057&	2003.441&	0.009\\
S64&	15.952&	3.947&	0.347&	0.161&	113.789&2.406&	154.985&31.883&	165.699&7.047&	2005.906&	6.192\\
S71&	39.052&	1.266&	0.916&	0.043&	67.151&	4.354&	336.842&2.120&	35.466&	2.578&	1689.433&	18.447\\
S175&	29.808&	0.001&	0.999&	0.001&	93.793&	0.001&	65.260&	0.001&	349.733&0.001&	2009.976&	0.001\\
\hline \hline
\end{tabular}
\label{elementsblack}
\end{center}
Following the stellar designation in column (1) we list consecutively the following 
quantities with their uncertainties:
semi-major axis, ellipticity, inclination, argument of periapse, longitude of ascending node, 
and the time of closest approach.
\end{table}

\begin{table}[htb]
\caption{Orbital Elements for Stars in the Red Disk}
\begin{center}
\begin{tabular}{rrrrrrrrrrrrrrrrrrrrrrccccc}
\hline \hline
1. & 2. & 3. & 4. & 5. & 6. &  7. &  8. &  9. &  10. & 11. &  12.&   13. \\ \hline 
Star & a & $\Delta$a & e & $\Delta$e & i & $\Delta$i   &$\omega$ & $\Delta$$\omega$ & $\Omega$ & $\Delta$$\Omega$ & t$_{clos}$ & $\Delta$t$_{clos}$ \\
     &[mpc]& [mpc]     &   &           &[deg]&[deg]&[deg] &[deg]          &[deg]  &[deg]          &   [yr]    &   [yr]   \\
\hline
S4&	14.555&	0.034&	0.443&	0.014&	80.386&	0.229&	286.823&0.229&	259.092&0.229&	1954.476&	0.011\\
S6&	25.229&	0.574&	0.891&	0.021&	86.459&	1.490&	119.175&0.974&	86.116&	3.782&	1932.803&	5.140\\
S14&	9.037&	2.426&	0.798&	0.287&	107.716&21.944&	378.668&28.904&	231.532&19.194&	2000.453&	3.942\\
S18&	9.253&	0.212&	0.461&	0.017&	111.727&4.011&	374.084&3.724&	54.202&	1.833&	1997.061&	0.006\\
S21&	8.662&	0.162&	0.772&	0.016&	59.530&	1.891&	161.173	&3.266&	262.930&0.917&	2027.290&	0.017\\
S22&	52.357&	2.553&	0.489&	0.062&	106.914&0.859&	94.366&	15.756&	289.859&3.953&	1996.959&	5.234\\
S23&	10.389&	1.945&	0.462&	0.205&	47.326&	6.303&	30.882&	13.980&	249.638&26.986&	2024.577&	8.064\\
S33&	31.326&	4.263&	0.664&	0.059&	64.057&	1.891&	304.183&2.464&	107.086&4.412&	1923.847&	11.286\\
S38&	5.598&	0.205&	0.812&	0.050&	157.774&15.011&	11.001&	9.626&	96.375&	8.308&	2003.406&	0.339\\
S54&	48.225&	10.890&	0.897&	0.018&	58.384&	2.120&	151.891&4.641&	254.164&5.672&	2002.326&	0.042\\
S66&	61.777&	2.985&	0.160&	0.028&	126.738&1.432&	144.271	&7.850&	87.892&	2.292&	1794.519&	13.396\\
S67&	47.708&	1.938&	0.082&	0.045&	131.895&2.292&	226.548&4.469&	79.756&	5.042&  1740.000&	15.852\\
S83&	58.717&	4.729&	0.377&	0.048&	125.592	&1.261&	207.697&7.391&	87.433&	7.506&	2049.789&	14.833\\
S85&	184.115&3.611&	0.773&	0.006&	85.084	&1.089&	157.907	&4.183&	107.544&0.974&	1930.384&	8.658\\
S87&	109.645&1.066&	0.163&	0.060&	117.514	&1.662&	334.779&3.610&	105.367&2.578&	~627.690&      12.927\\
S89&	42.801&	2.027&	0.651&	0.224&	91.731	&1.490&	123.644	&1.089&	234.282&1.547&	1777.211&	21.179\\
S91&	78.892&	1.958&	0.322&	0.034&	113.560	&2.005&	366.120	&4.870&	101.643&2.636&	1086.879&	21.025\\
S96&	54.529&	1.070&	0.289&	0.078&	127.712	&2.865&	238.179	&5.730&	121.582&4.927&	1688.413&	22.780\\
S97&	92.859&	2.783&	0.382&	0.033&	112.300	&1.891&	38.503	&4.354&	109.148&2.177&	2161.556&	14.970\\
S145&	42.278&	0.501&	0.550&	0.016&	83.136	&7.506&	177.388	&2.406&	263.904&0.229&	1808.606&	4.260\\
\hline \hline
\end{tabular}
\label{elementsred}
\end{center}
Following the stellar designation in column (1) we list consecutively the following 
quantities with their uncertainties:
semi-major  axis, ellipticity, inclination, argument of periapse, longitude of ascending node, 
and the time of closest approach.
\end{table}

\begin{table}[htb]
\caption{Position Angles of the Black Disk}
\begin{center}
\begin{tabular}{rrrrrrrrrrrrrrrrrrrrrrccccc}
\hline \hline
1. & 2. & 3. & 4. & 5. & 6. &  7.  \\ \hline 
Star & m1(t,R.A.) & $\Delta$m1 & m2(t,Dec.) & $\Delta$m2 & $\Phi$ & $\Delta$$\Phi$   \\
     &[mas/yr]& [mas/yr]     & [mas/yr]  & [mas/yr]           &[deg]&[deg]  \\
\hline
S1&	0.767&	0.086&	8.710&	0.086&	5.035	& 0.562\\
S2&	1.608&	0.951&	-21.019&0.951&	175.624	& 2.584\\
S8&	-14.616&0.233&	14.989&	0.233&	-44.278	& 0.637\\
S9&	-10.771	&0.472&	22.951&	0.472&	-25.141	& 1.066\\
S12&	4.253	&0.292&	17.785&	0.292&	13.450	& 0.914\\
S13&	0.678	&0.334&	-17.274&0.334&	177.751	& 1.108\\
S17&	-5.771	&0.319&	-22.271&0.319&	-165.473 &0.795\\
S19&	-5.404	&0.241&	12.290&	0.241&	-23.735	& 1.030\\
S24&	2.820	&0.068&	13.475&	0.068&	11.820	 &0.284\\
S29&	-1.500	&0.054&	10.578&	0.054&	-8.070	 &0.289\\
S31&	-6.491	&0.159&	12.484&	0.159&	-27.471	 &0.646\\
S39&	4.367	&0.195&	-12.792&0.195&	161.153	 &0.826\\
S42&	7.952	&0.076&	13.161&	0.076&	31.141	 &0.282\\
S55&	-4.814	&1.328&	28.810&	1.328&	-9.487	 &2.605\\
S60&	-2.209	&0.163&	16.687&	0.163&	-7.542	 &0.556\\
S62&	15.992	&1.793&	-20.849&1.793&	142.511	 &3.910\\
S64&	3.155	&0.173&	-15.407&0.173&	168.428	 &0.630\\
S71&	-5.393	&0.049&	-9.191&	0.049&	-149.594 &0.264\\
S175&	1.688	&0.079&	-5.430&	0.079&	162.729	 &0.7912\\
\hline \hline
\end{tabular}
\label{pablack}
\end{center}
Following the stellar designation in column (1) we list consecutively the following 
quantities with their uncertainties:
slopes of the R.A. and Dec. data as a function of time
and the corresponding position angles $\phi$.
\end{table}

\begin{table}[htb]
\caption{Position Angles of the Red Disk}
\begin{center}
\begin{tabular}{rrrrrrrrrrrrrrrrrrrrrrccccc}
\hline \hline
1. & 2. & 3. & 4. & 5. & 6. &  7.  \\ \hline 
Star & m1(t,R.A.) & $\Delta$m1 & m2(t,Dec.) & $\Delta$m2 & $\Phi$ & $\Delta$$\Phi$   \\
     &[mas/yr]& [mas/yr]     & [mas/yr]  & [mas/yr]           &[deg]&[deg]  \\
\hline
S4 &	-22.725 &0.305	 &-4.321 &	0.305 &	-100.766 &0.755\\
S6 &	-14.624	 &0.161	 &-1.734 &	0.161 &	-96.761	  & 0.628\\
S14 &	-17.048	 &0.510	 &-15.601 &	0.510 &	-132.461 &1.265\\
S18 &	18.905	 &0.443	 &13.206 &	0.443 &	55.064	  & 1.101\\
S21 &	26.326	 &0 .557 &5.246	  &      0.557 &78.730	   & 1.189\\
S22 &	-5.898	 &0.024	 &1.998	 &      0.024 &	-71.287	    &0.219\\
S23  &	22.507	 &0.392	 &2.768    &	0.392 &	82.990	   &0.990 \\
S33 &	15.026	 &0.100	 &-1.445 &	0.100 &	95.493	    &0.378\\
S38 &	29.086	 &0.919	 &1.432	   &      0.919 &87.182	   &1.809\\
S54 &	11.014	 &0.048 &5.224    &	0.048 &	64.624	  & 0.228\\
S66 &	-9.965	 &0.028	 &-0.174 &	0.028 &	-91.001	   &0.161\\
S67 &	10.856	 &0.044	 &0.054	   &      0.044 &89.715	   &0.232\\
S83 &	9.884	 &0.029	 &2.013	    &    0.029 &78.486	   &0.166\\
S85 &	4.688	 &0.005	 &-1.380 &	0.005 &	106.407	   &0.058\\
S87 &	-5.897	 &0.010	 &2.186	   &     0.010 &-69.664	   &0.094\\
S89 &	11.564	 &0.067	 &8.097	  &      0.067 &55.001	   &0.274\\
S91 &	-7.563	 &0.017	 &0.848	   &     0.027 &-83.604	   &0.128\\
S96 &	8.376	 &0.026	 &-2.054 &	0.026 &	103.781	   &0.174\\
S97 &	-7.486	 &0.013	 &1.026	  &     0.013 &	-82.195	   &0.101\\
S145 &	11.179	 &0.046	 &1.170	   &  0.046 &	84.023	   &0.235\\
\hline \hline
\end{tabular}
\label{pared}
\end{center}
Following the stellar designation in column (1) we list consecutively the following 
quantities with their uncertainties:
slopes of the R.A. and Dec. data as a function of time
and the corresponding position angles $\phi$.
\end{table}

\begin{table}[htb]
\caption{Position Angles of the Linear Stellar Trajectories}
\begin{center}
\begin{tabular}{rrrrrrrrrrrrrrrrrrrrrrccccc}
\hline \hline
1. & 2. & 3. & 4. & 5. & 6. &  7.  \\ \hline 
Star & m1(t,R.A.) & $\Delta$m1 & m2(t,Dec.) & $\Delta$m2 & $\Phi$ & $\Delta$$\Phi$   \\
     &[mas/yr]& [mas/yr]     & [mas/yr]  & [mas/yr]           &[deg]&[deg]  \\
\hline
S5&	-6.121&	0.207&	7.610&	0.318&	-38.813&	1.502\\
S7&	-3.768&	0.063&	-1.826&	0.117&	-115.849& 1.494\\
S10&	-4.823&	0.090&	3.667&	0.059&	-52.752	&0.677\\
S11&	8.486&	0.152&	-4.849&	0.241&	119.743&1.303\\
S20&	-4.661&	0.203&	-5.363&	0.214&	-139.005& 1.675\\
S25&	-2.441&	0.105&	1.724&	0.181&	-54.768	&3.063\\
S26&	5.700&	0.131&	1.930&	0.157&	71.292	&1.475\\
S27&	0.215&	0.143&	3.609&	0.173&	3.416	&2.261\\
S28&	4.381&	0.412&	5.065&	0.392&	40.860	&3.452\\
S30&	0.318&	0.102&	3.296&	0.098&	5.504	&1.757\\
S32&	-3.609&	0.125&	-0.199&	0.238&	-93.150	&3.774\\
S34&	9.899&	0.209&	4.441&	0.156&	65.837	&0.876\\
S35&	1.834&	0.097&	3.727&	0.187&	26.197	&1.656\\
S36&	0.268&	0.246&	-1.360&	0.431&	168.848	&10.561\\
S37&	-6.324&	0.351&	9.605&	0.283&	-33.359	&1.653\\
S40&	4.172&	0.585&	5.165&	0.935&	38.929	&6.414\\
S41&	1.331&	0.130&	-3.197&	0.182&	157.405	&2.302\\
S43&	5.119&	0.177&	6.135&	0.430&	39.839	&2.201\\
S44&	-6.662&	0.559&	-8.450&	0.589&	-141.746& 3.038\\
S45&	-5.688&	0.162&	-4.037&	0.117&	-125.363& 1.100\\
S46&	0.966&	0.186&	4.566&	0.161&	11.950	&2.268\\
S47&	-3.058&	0.448&	2.789&	0.186&	-47.633	&4.594\\
S48&	-1.626&	0.212&	10.040&	0.418&	-9.198	&1.238\\
S49&	15.222&	0.268&	-0.760&	0.664&	92.859	&2.494\\
S50&	-1.370&	0.362&	10.459&	0.327&	-7.462	&1.963\\
S51&	8.422&	0.509&	7.655&	0.397&	47.730	&2.273\\
S52&	4.627&	0.501&	-5.721&	0.298&	141.033	&3.369\\
S53&	7.096&	0.366&	9.465&	0.504&	36.860	&2.039\\
S56&  -18.748&	0.685&	-1.319&	0.411&	-94.026	&1.259\\
S57&	-9.770&	0.521&	-0.312&	0.360&	-91.828	&2.112\\
S58&	7.686&	0.356&	5.449&	0.202&	54.667	&1.603\\
S59&	7.458&	0.375&	-1.606&	0.342&	102.154	&2.579\\
S61&	-4.487&	0.561&	-6.718&	1.017&	-146.258& 5.195\\
S63&	-13.15&	0.847&	4.335&	0.549&	-71.755	&2.419\\
S65&	2.401&	0.097&	-1.616&	0.124&	123.940	&2.305\\
S68&	3.971&	0.236&	3.108&	0.148&	51.946	&2.119\\
S69&	-1.786&	0.207&	2.052&	0.558&	-41.037	&8.384\\
S70&	-4.141&	0.235&	-3.600&	0.263&	-131.006& 2.626\\
S72&	9.101&	0.200&	-5.645&	0.133&	121.811	&0.825\\
S73&	-9.223&	0.245&	-7.771&	0.156&	-130.115& 0.941\\
S74&	-0.170&	0.209&	5.026&	0.242&	-1.941	&2.380\\
\hline \hline
\end{tabular}
\label{linear1}
\end{center}
Following the stellar designation in column (1) we list consecutively the following 
quantities with their uncertainties:
slopes of the R.A. and Dec. data as a function of time
and the corresponding position angles $\phi$.
\end{table}

\begin{table}[htb]
\caption{Position Angles of the Linear Stellar Trajectories}
\begin{center}
\begin{tabular}{rrrrrrrrrrrrrrrrrrrrrrccccc}
\hline \hline
1. & 2. & 3. & 4. & 5. & 6. &  7.  \\ \hline 
Star & m1(t,R.A.) & $\Delta$m1 & m2(t,Dec.) & $\Delta$m2 & $\Phi$ & $\Delta$$\Phi$   \\
     &[mas/yr]& [mas/yr]     & [mas/yr]  & [mas/yr]           &[deg]&[deg]  \\
\hline
S75&	7.138&	0.249&	2.330&	0.321&	71.921	&2.404\\
S76&	-3.329&	0.182&	4.898&	0.161&	-34.201	&1.700\\
S77&	9.536&	0.439&	-6.606&	0.524&	124.711	&2.460\\
S78&   -16.728&	0.429&	-5.989&	0.360&	-109.697& 1.190\\
S79&	0.040&	0.218&	4.269&	0.449&	0.536	&2.932\\
S80&	-4.640&	0.221&	6.325&	0.435&	-36.261	&2.288\\
S81&	5.072&	1.908&	6.529&	0.739&	37.842	&10.906\\
S82&	-8.689&	0.329&	-14.942&0.373&	-149.821& 1.128\\
S84&	3.926&	0.084&	1.282&	0.218&	71.916	&2.903\\
S86&	-0.892&	0.308&	-4.872&	0.392&	-169.627& 3.601\\
S88&	-3.941&	0.265&	-7.715&	0.227&	-152.939& 1.702\\
S90&	1.713&	0.349&	1.266&	0.228&	53.536	&7.452\\
S92&	5.530&	0.095&	2.238&	0.303&	67.966	&2.718\\
S93&	-2.972&	0.349&	-1.755&	0.374&	-120.553& 6.099\\
S94&   -13.564&	0.513&	2.089&	0.735&	-81.243	&3.052\\
S95&	5.273&	0.295&	0.987&	0.101&	79.402	&1.208\\
S98&	-7.985&	0.186&	1.922&	0.284&	-76.465	&1.947\\
S99&   -10.093&	0.459&	0.240&	0.336&	-88.638	&1.906\\
S100&	-1.392&	0.202&	-2.557&	0.200&	-151.432& 3.969\\
S101&	2.168&	0.422&	5.836&	0.400&	20.376	&3.864\\
S102&	-5.376&	0.594&	8.018&	0.512&	-33.842	&3.380\\
S103&	11.849&	0.405&	-3.727&	0.736&	107.463	&3.288\\
S104&	10.255&	0.344&	-2.270&	0.507&	102.481	&2.733\\
S105&	3.908&	0.336&	-7.620&	0.482&	152.849	&2.484\\
S106&	1.135&	0.275&	-0.981&	0.570&	130.845	&17.841\\
S107&	-0.580&	0.112&	4.581&	0.228&	-7.217	&1.426\\
S108&	3.868&	0.420&	0.449&	0.422&	83.377	&6.208\\
S109&	7.021&	0.374&	-4.926&	0.278&	125.050	&2.089\\
S110&	-2.956&	0.267&	-1.221&	0.251&	-112.442& 4.534\\
S111&	-2.490&	0.245&	-7.337&	0.255&	-161.253& 1.819\\
S112&	3.062&	0.439&	10.822&	0.341&	15.797	&2.205\\
S146&	-3.465&	0.905&	-0.189&	0.986&	-93.118	&16.281\\
\hline \hline
\end{tabular}
\label{linear2}
\end{center}
Continuation of table:
Following the stellar designation in column (1) we list consecutively the following 
quantities with their uncertainties:
slopes of the R.A. and Dec. data as a function of time
and the corresponding position angles $\phi$.
\end{table}

\end{document}